\documentclass[11pt]{article}

\usepackage{subfig}
\usepackage{amsmath}
\usepackage{authblk}
\usepackage{a4wide}
\usepackage{soul}

\usepackage[utf8]{inputenc}
\usepackage[T1]{fontenc}

\title{On constant-time quantum annealing and guaranteed approximations for graph optimization problems} %TODO Please add

\author[1]{Arthur Braida} %{Atos quantum lab, France \and Laboratoire d'Informatique Fondamentale d'Orléans, France }{arthur.braida@atos.net}{https://orcid.org/0000-0002-1825-0097}{}
\author[2]{Simon Martiel} %{Atos quantum lab, France}{simon.martiel@atos.net}{}{}
\author[3]{Ioan Todinca} %{Laboratoire d'Informatique Fondamentale d'Orléans, France}{ioan.todinca@univ-orleans.fr}{}{}
\affil[1]{Atos quantum lab and Laboratoire d'Informatique Fondamentale d'Orléans, France, \texttt{arthur.braida@atos.net}, https://orcid.org/0000-0002-1825-0097}
\affil[2]{Atos quantum lab, France, \texttt{simon.martiel@atos.net}}
\affil[3]{Laboratoire d'Informatique Fondamentale d'Orléans, France, \texttt{ioan.todinca@univ-orleans.fr}}

\usepackage{tikz}
\usepackage{bbold}
\usepackage{caption}
\usepackage{comment}

\newtheorem{proposition}{Proposition}
\newtheorem{corollary}{Corollary}
\newtheorem{remark}{Remark}
\newtheorem{claim}{Claim}

\newcommand{\ket}[1]{|#1\rangle}
\newcommand{\bra}[1]{\langle #1|}

\begin{document}

\maketitle

\begin{abstract}
Quantum Annealing (QA) is a computational framework where a quantum system's continuous evolution is used to find the global minimum of an objective function over an unstructured search space. It can be seen as a general metaheuristic for optimization problems, including NP-hard ones if we allow an exponentially large running time. While QA is widely studied from a heuristic point of view, little is known about theoretical guarantees on the quality of the solutions obtained in polynomial time.

In this paper we use a technique borrowed from theoretical physics, the Lieb-Robinson (LR) bound, and develop new tools proving that short, constant time quantum annealing guarantees constant factor approximations ratios for some optimization problems when restricted to bounded degree graphs. Informally, on bounded degree graphs the LR bound allows us to retrieve a (relaxed) locality argument, through which the approximation ratio can be deduced by studying subgraphs of bounded radius.

We illustrate our tools on problems MaxCut and Maximum Independent Set for cubic graphs, providing explicit approximation ratios and the runtimes needed to obtain them. Our results are of similar flavor to the well-known ones obtained in the different but related QAOA (quantum optimization algorithms) framework. 

Eventually, we discuss theoretical and experimental arguments for further improvements.

%Quantum annealing is a general setting to solve combinatorial problems. Using a version of the Lieb-Robinson bound to recover a notion of locality, we develop a new method to approximate problems in finite time. It uses the combination of the Hamiltonian locality and the speed limit in the information flow in quantum annealing to derive bounds on the approximation ratio.  We apply it to MaxCut and Maximum Independent Set on regular graphs to give the runtime needed to get an approximation ratio. We then develop an algorithm to find the best approximation for MaxCut ratio in short time. In practice, we compute it only for relatively small graphs and give insights about how quantum annealing works on MaxCut to argue about worst cases.

\end{abstract}

\vfill 

\thispagestyle{empty}

\pagebreak
\setcounter{page}{1}

\section{Introduction}
\label{sec:intro}

Quantum Annealing (QA)\cite{APOLLONI1989, Kadowaki_1998, farhi2000quantumoriginal} is a computational framework where a quantum system's continuous evolution is used to find the global minimum of an objective function over an unstructured search space. This metaheuristic proceeds by preparing a simple, problem independent, 
quantum state and evolving it along a problem-dependent trajectory, reaching a final state that overlaps non trivially with the optimal solution. A well known theoretical result, called the adiabatic theorem, ensures that this overlap will be large if we respect some simple condition on the initial quantum state and if the system's evolution is slow enough (adiabaticity condition). This time-continuous approach for quantum computing can be shown to be equivalent to the more traditional, circuit based quantum computing model \cite{farhi2000quantumoriginal, Albash_2018}, in the sense that quantum circuits can simulate QA using a number of gates that only depends polynomially on the precision of the simulation. Conversely, quantum states produced by quantum circuits can be prepared via a polynomial time QA (i.e. a QA trajectory that has polynomially small spectral gap). In \cite{Albash_2018}, the authors provide a complete review of this equivalence and generalize some known results of digital quantum computing to the continuous setting, such as
%Several reasons make quantum annealing a very appealing algorithmic approach for combinatorial optimization problems. In the most up-to-date review of QA \cite{Albash_2018} by Albash and Lidar, they state that this framework is equivalent to digitized quantum computing setting. 
  quadratic speed-up for search in an unstructured data-set, thus generalizing Grover's algorithm \cite{grover96}. The QA framework has been (mostly numerically) investigated~\cite{Farhi_2001, Childs2002} as a tool to solve combinatorial optimization problems including $k$-clique and Exact Cover. Moreover, many classical NP-hard optimization problems can be expressed with reasonable efforts in the QA framework~\cite{lucas14}, making it a very general heuristic. On the negative side, the adiabaticity condition asks for a "slow enough" evolution in order to obtain the optimal solution. In general, QA needs exponential time for NP-hard problems~\cite{altshuler2010adiabatic}. While there is not much hope for finding exact solutions efficiently, one can still try and study QA as a probabilistic approximation algorithm and quantify the achievable approximation ratio for various combinatorial problems when run in reasonable time.

We believe that quantum annealing is a promising, general framework for designing approximation algorithms, but to the best of our knowledge there are no general results on guaranteed approximation bounds in the literature. The main obstacle seems to be the current lack of mathematical tools for analyzing the solutions obtained in limited time. The purpose of this work is to make a step forward toward such tools, and show that QA produces reasonable solutions for specific graph optimization problems, on specific graph classes.  
%\arthur{To the best of our knowledge, no general results on approximation's guarantees exist.}  
%

%
In a closed system, the evolution of a quantum state $\ket{\psi(t)}$ is guided by Schr\"odinger's equation:

\begin{align}
\label{eq:shrod}
i \frac{\partial}{\partial t}|\psi(t)\rangle=H(t) |\psi(t)\rangle \quad \text{ for } t \in [0,T]
\end{align}

where $T$ is the runtime of the evolution, or in our case, of the algorithm (taking the constant $\hbar$ as unity) and $H(t)$ is the instantaneous Hamiltonian of the system.
%In fact, one needs to be careful in normalizing with the largest interaction term norm which in our case is in $\mathcal{O}(1)$.\ioan{Je pense qu'on risque de nous reprocher ce point. Ignorer une constante multiplicative n'est pas trop gênant en algorithmique... sauf lorsque la constante est extrême, or ici c'est le cas. La dernière phrase est un avertissement mais il n'est pas clair si nous-mêmes avons fait gaffe. Je radote mais je n'ao pas de proposition plus constructive...)} \arthur{ ok après avoir discuté avec Thomas, notre cher physicien, il faut voir $\Tilde{H}=\frac{H}{\hbar}$ et l'ordre de grandeur des termes de $\Tilde{H}$ est donné en MHz ou GHz car $E=h \nu $, ça dépend des techno utilisé. Cela revient à des temps en micro ou nano secondes donc temps constant raisonnable, il y a un facteur $2\pi$ entre notre temps $T$ et la fréquence de la techno.} 
%The quantum state at time $t$ is written $|\psi(t)\rangle$ \cite{nielsen2011quantum}, and $H(t)$ gives the instantaneous Hamiltonian of the problem. 
QA can be used for computation (see., e.g., \cite{Albash_2018} and Section~\ref{ssec:prelim} for more details) by clever choices of $H$. Typically, we choose a problem-independent initial Hamiltonian $H_0 = H(0)$, and a target Hamiltonian $H_1 = H(T)$ whose ground state (i.e., the eigenvector of minimum eigenvalue) corresponds to an optimum solution of our given problem. The time-dependent Hamiltonian $H(t)$ will then follow a specific trajectory according to a chosen schedule. 
%This schedule will only have an impact on the speed of convergence.

The adiabatic theorem ensures that, under some technical assumptions, if time $T$ is large enough (exponentially large), the system evolves towards the ground state of $H_1$, hence towards an optimal solution. No general result is known for polynomial time or even shorter, constant time.
In this work we focus on short, constant-time evolution, and we are interested in getting an approximate solution for graph optimization problems. Indeed, we would like to understand what guarantee on the results quantum computers can expect to reach while they are stable only for a very short time. 

\subparagraph{Related work.}
In the past years, a constant depth digital variant of QA, called Quantum Approximate Optimization Algorithm (QAOA) has been intensively studied as a promising heuristic for combinatorial optimization. In consists in a discretization of a QA trajectory followed by a parametrization of the time steps of this discretization. Some approximation bounds have been derived in this model for particular classes of instances of MaxCut~\cite{farhi2014quantum} (among other similar problems). 
More recently \cite{farhi2020quantum}, it has been shown that for some problems, QAOA requires at least $O(\log{n})$ depth in order to have a shot at beating Goesman-Williamson ratio for MaxCut problem. This result has been transposed \cite{moosavian2021limits} to QA using a technique borrowed from theoretical physics called the Lieb-Robinson bound that provides a tool to bound correlation strength in continuous time dynamics. This same tool was also used to improve simulation schemes for geometrically constrainted continuous dynamics \cite{Haah_2021, Tran_2019}.

\subparagraph{Our contribution.}
The main technical contribution in this work is the use of LR-like bounds to retrieve a relaxed locality argument in QA.

We see QA as a probabilistic algorithm and analyze the expected cost of its output. For the graph optimization problems considered here, the output cost can be decomposed as the sum of costs of vertices and edges. Hence the challenge will be to evaluate the expected values of these terms, called energies, for vertices and edges.
Informally speaking, when considering a Hamiltonian whose interaction terms are restricted to the input graph, by making the quantum system evolve according to this Hamiltonian and observing the energy of an edge or a vertex, the result will be comparable to the same observation made on a quantum system evolving according to the Hamiltonian restricted to a ball of bounded radius around the observed location. LR-type bounds give us a tool to effectively bound the difference between these two quantities.
%local Hamiltonian (e.g. whose interaction terms are restricted to some graph), any local observation of a quantum system evolving according to this Hamiltonian (e.g. the energy of an edge or a vertex) will be comparable to the same observation made on a quantum system evolving according to the Hamiltonian restricted to a ball of bounded radius around the observed location. The LR bound gives us a tool to effectively bound the difference between these two quantities.
%Informally speaking the LR bound tells us that looking at a local Hamiltonian, focused on an edge for example, and if the system evolves for a short amount of time, then the evolution is very close to the same evolution but only on a neighborhood of the edge in the input graph. 

With this relaxed notion of locality, we are able to reuse the same combinatorial arguments as developed in the QAOA framework~\cite{farhi2014quantum}. This general method is firstly applied to MaxCut, which is a standard benchmark for QAOA. In order to illustrate the generality of our method, we also apply it to the Maximum Independent Set (MIS) problem. We show that we reach approximation ratios better than random guess in these two cases, for regular (cubic) graphs. We guarantee ratios of 0.5933 for MaxCut and of 0.3171 for MIS, to be compared with the ratios obtained by naive algorithms, of 0.5 and 0.25 respectively. 
%The improvement may not be impressive\simon{ce n'est pas à nous de juger de ça, non?}, but the key message is that LR-type bounds are good enough (of order 0.01 on balls of radius one)\simon{pas compris ça?} to provide guaranties for the approximation ratios of these problems\marginpar{\ioan{Hm, ça laisse l'impression qu'on ne fait que quelques petits calculs}}\marginpar{\simon{Je serai pour dropper la partie sur les cycles pour ne pas noyer notre résultat}}.  Furthermore, we develop the intuition on how QA effectively works when solving MaxCut by computing the ratio of specific graphs that we argue to be the "worst" cases. 
We perform further numerical studies suggesting that constant time QA can achieve an approximation ratio of $0.6963$ for MaxCut on 3-regular graphs, which is significantly better than the one we actually prove. 
%To illustrate the locality in MaxCut, we detail in the appendices the much simpler case of the 2-regular graphs to observe the behaviour of short time QA.\marginpar{\ioan{Pas sûr que la partie sur les cycles mérite d'être mentionnée}} %Unsurprisingly, our approximation ratios are not as good as the ones of QAOA, since we did not optimize the trajectory of the evolution --- the analog of the parameters optimization in step (2) of QAOA. There are two ways to improve the ratio obtained by QA. One can smartly change the trajectory follow by the evolution and it can also help to find a good perturbative Hamiltonian that boost the evolution. We will discuss some experimental improvements obtained by the second method.
%\ioan{On rame un peu pour dire que ce qu'on fait est bien mais que le résultat n'est pas aussi bon que chez les autres, puisque ce n'est qu'une première tentative, mais je ne vois pas comment échapper à cette réalité.} \simon{Je suis d'accord. Je pense qu'à ce stade on peut se contenter de dire qu'on va donner des pistes sur comment améliorer le ratio en explorant divers Hamiltoniens, sans trop s'étendre.}

\subparagraph{Organization of the paper.}
The paper organization is as follows. In Section~\ref{sec:loc} we give the necessary notations and detail the setting in which we use QA. We then develop the notion of local algorithm and explain why focusing on regular (in particular cubic) graphs is convenient when dealing with such algorithms. We then show the main tool to recover locality in quantum annealing, namely a Lieb-Robinson bound. In Section~\ref{sec:app}, we apply the method described to MaxCut and MIS to output an approximation ratio at a fixed time, on cubic graphs. Section~\ref{sec:exp} brings 
%an algorithm to output the best approximation ratio for short runtime. 
arguments suggesting that the actual approximation ratio of short-time QA for MaxCut in cubic graphs is significantly better than the one that is proved in the previous section.
In practice, we are only able to compute the ratios for small cubic graphs (up to 30 nodes) and get the ratio up to $T=5$. Eventually, we talk about the non-tightness of the LR bound used here and where we can earn to improve the ratio. 

\section{Locality in short-time QA}
\label{sec:loc}

%\ioan{[Je vois dans la version actuelle plusieurs sources de "sloppyness", d'ambiguïté, dans cette section. (1) La fonction $C$ est définie sur $\{0, 1\}^n$ mais en fait elle est associée à un graphe. On peut laisser comme ça. (2) Dans la définition de $H_1$ on a des $N_i$ et des $M_{ij}$ sans trop dire à quoi ça correspond mathématiquement. (3) $U^G_{s,t}$ n'est pas défini.]}

%\simon{
%(1) On peut définir $C$ sur $\{0, 1\}^V$ (2) %On peut en effet tenter de les définir %proprement en fonction de $v(\cdot)$ et %$u(\cdot, \cdot)$. (3) En effet :)
%}

\subsection{Preliminaries}\label{ssec:prelim}

\subparagraph{Optimization problems on graphs.}
We consider optimization problems whose input is a graph $G=(V,E)$ and the output is a set of vertices, or a two-partition of the vertex set, maximizing some global cost function. We are interested in combinatorial graph problems where the global cost is a sum of cost functions localized on nodes and edges of the input graph. 

Formally, we are given as input graph $G=(V,E)$ on $n$ vertices, and our cost function $C$ associates to any  $n$-bit vector $x$ (bit $x_i$ corresponding to the boolean value of node $i$) a value  $C(x)=\sum_{i\in V(G)} v(x_i) + \sum_{\langle i,j \rangle \in E(G)} u(x_i,x_j)$, for some functions $v:\{0,1\}  \rightarrow \mathbb{N}$ and $u: \{0,1\} \times \{0,1\}\rightarrow \mathbb{N}$.

%Thus, we write $C(x)=\sum_{i\in V(G)} v_i(x) + \sum_{\langle i,j \rangle \in E(G)} u_{ij}(x)$ for some graph $G$. In our case, $n$ is also the number of nodes of $G$, and the $n$-bit vector $x$ associates a bit $x_i$ to each node $i$ of $G$, hence $x$ will represent a subset or a two-partition of the vertex set. In general, the local cost function $v_i(x)$ depends only on $x_i$ the $i^{th}$ bit of $x$ and it is the same for every node, $u_{ij}$ depends only on $(x_i,x_j)$ and is the same for every edge.

The optimization problem consists in finding a bitstring $x$ maximizing value $C_G(x)$. Bitstring $x$ can represent a subset of vertices (e.g., in the case of Maximum Independent Set) or a two-partition of the vertex set (e.g., in the case of MaxCut). 

%In order to run a quantum annealing, $C$ must be first encoded into an Hamiltonian acting over a $n$-qubit system (see e.g. the classical book~\cite{nielsen2011quantum} for notations and basic notions in quantum computing). In other words, we need to construct a diagonal hermitian operator $H_1\in Herm(\mathbb{C}^{2^n})$ such that for all $x \in\{0,1\}^n$, $H_1 |x\rangle= - C(x) |x\rangle$. This construction is achieved by encoding each clause of $C$ separately: $H_1 = -\sum_{i\in V(G)} N_i - \sum_{\langle i,j \rangle \in E(G)} M_{ij}$.
%Thus, the structure of $C$ is preserved by the final Hamiltonian $H_1$ and, local clauses become local terms (also called observables). Here, $N_i$ encodes the local cost function of the node $v_i$ and $M_{ij}$ the local cost function of the edge $u_{ij}$. 
%A general expression is $H_1=-\sum_i O_i - \sum_{\langle i,j \rangle}O_{ij}$ that we want to minimize because QA by nature tries follow the state of minimum energy. So, for each $x \in \{0,1\}^n$, the quantum state associated $|x\rangle$ respects $H_1 |x\rangle=C(x) |x\rangle$, i.e. a pair $(x,C(x))$ is an eigenvector/eigenvalue pair of $H_1$. 

\subparagraph{Quantum annealing and combinatorial optimization.} 
For notations and basic notions in quantum computing, we refer to the classical book of Nielsen and Chuang~\cite{nielsen2011quantum}. 

A quantum annealing algorithm for optimizing a cost function $C$ on $n$-vertex graphs is defined by three ingredients.
\begin{enumerate}
\item A problem-independent initial Hamiltonian $H_0$ (i.e., hermitian operator in $Herm(\mathbb{C}^{2^n})$) to start our annealing. A standard choice for this is $H_0 = - \sum_{i} X_{i}$ where $X_{i}$ is the bit-flip operator acting on qubit $i$. The ground state (i.e. the eigenvector of minimum eigenvalue) of this Hamiltonian is the product state $\ket{\psi_0} = \left(\frac{\ket{0}+\ket{1}}{\sqrt{2}}\right)^{\otimes n}$,  the uniform superposition of all possible bit-strings of length $n$. This state is particularly easy to prepare (it is not entangled) and will be the starting point of our computation.
\item A target Hamiltonian $H_1$ encoding the cost function $C$, in the sense that for all $x \in\{0,1\}^n$, $H_1 |x\rangle= - C(x) |x\rangle$. This completely defines $H_1$ and thus is diagonal in the computational basis. This construction is achieved by encoding each clause of $C$ separately: $H_1 = -\sum_{i\in V(G)} N^{(i)} - \sum_{\langle i,j \rangle \in E(G)} M^{(i,j)}$.
Thus, the structure of $C$ is preserved by the final Hamiltonian $H_1$ and, local clauses become local terms (also called observables). Here, $N^{(i)}$ encodes the local cost function $v(x_i)$ of node $i$ acting on qubit $i$ and $M^{(i,j)}$ the local cost function $u(x_i, x_j)$ of edge $\langle i,j \rangle$ acting on qubits $i$ and $j$. They can be defined as follow:
\begin{align*}
    &N^{(i)} = (v(0)\ket{0}\bra{0} + v(1) \ket{1}\bra{1})\otimes \mathbb{I}_{V-\{i\}} \\
    &M^{(i,j)} = (u(0,0)\ket{00}\bra{00} + u(0,1) \ket{01}\bra{01} + u(1,0) \ket{10}\bra{10} + u(1,1) \ket{11}\bra{11})\otimes \mathbb{I}_{V-\{i,j\}}
\end{align*}
\item A running time $T$ and a trajectory $H(t)$ for $t \in [0,T]$ from $H_0 = H(0)$ to $H_1 = H(T)$. The instant Hamiltonian $H(t)$ will be defined here by the linear interpolation $H(t)=(1-\frac{t}{T})H_0+\frac{t}{T}H_1$, but one can imagine any smooth trajectory such that $H(0) = H_0$ and $H(T) = H_1$. 
\end{enumerate}

Driving our quantum system according to the Hamiltonian $H$ over a graph $G$ between times $s < t$ induces a unitary evolution $U^G_{s,t}$. Using this notation, the state of the quantum system after a time $t$ is given by $\ket{\psi^G(t)} = U^G_{0,t}\ket{\psi_0}$. This evolution operator is defined by $U^G_{0,0}=I$ and respects the Schrodinger equation where the total Hamiltonian $H$ depends on the graph $G$:

\begin{align}
\label{eq:shrod2}
i \frac{\partial}{\partial t}U^G_{0,t}=H(t) U^G_{0,t} \quad \text{ for } t \in [0,T]
\end{align}
In general, when running a QA to solve a combinatorial problem, the final state is a superposition of all bit-strings inducing a probability distribution over all possible solutions. By the adiabatic theorem, increasing the running time $T$ effectively concentrates this distribution around the optimal solution. In our case, we want to derive an approximation ratio, therefore we need to compute the expected value of the cost function over the final distribution. In quantum computing \cite{nielsen2011quantum}, this expected value is defined as $E_f(T)=-\langle \psi^G(T)|H_1|\psi^G(T) \rangle$. By linearity, this expected value can be written as: 
\begin{align}\label{eq:expanded_expected}
    E_f(T) = \sum_i \bra{\psi^G(T)}N^{(i)}\ket{\psi^G(T)} +\sum_{\langle i,j \rangle} \bra{\psi^G(T)} M^{(i,j)}\ket{\psi^G(T)}
\end{align}
Thus, we are ultimately interested in producing lower bounds for this quantity in the non-adiabatic regime where $T$ is small.

%This last expression is interesting because when dealing with local algorithms, we can reduce the computation of the final expected value to values computed on subgraphs formed by the neighborhoods, at bounded distance, around vertices and edges. \simon{Je pense que c'est pas super clair encore cette portion. Il faudrait avoir expliqué que les $O_i/O_{i,j}$ sont locaux wrt la topologie du graphe. De plus "local algorithm" est pas définit (même informellement?). J'ai presque envie de rajouter une subsection "Analyzing probabilistic local graph algorithms" un truc du genre.}\ioan{Cette histoire de voisinage à distance au plus $p$ autour d'un sommet ou d'une arête mérite une définition claire, car ce n'est pas standard en termes de graphes (ce n'est pas le voisinage induit, puisqu'on ne regarde que les arêtes incidentes au "centre", cf. $\Omega_0$. J'ai néanmoins remplacé "radius" par "distance" car "radius" a une autre signification -- Gutman et al. parlent bien de "distance" dans leur papier QAOA.}
%If the number of these subgraphs is finite and we can count them, then the final expected value can be computed. This is why we will focus on regular graphs, as in studies on QAOA.

\subsection{Locality and restriction to regular graph}

%\subparagraph{Locality.} 
We denote by $| \psi^G \rangle$  the state produced by the QA algorithm on the input graph $G$. This state depends on the whole input, in particular this entails that each term in Eq. \ref{eq:expanded_expected} might depend on the structure of the whole graph $G$.

Let us assume for a moment that our algorithm is "local" in the sense that, for each node and edge, it does not create correlations farther than a neighborhood at distance $p$. As a consequence, the expectation value of a local observable $N^{(i)}$/$M^{(i,j)}$ only depends on the structure of the neighborhood at distance $p$ around this vertex/edge. 
Here, by neighborhood at distance $p$ around subgraph $X$, we mean the subgraph $\Omega$ of $G$ containing $X$ and all vertices and edges situated on a path of $G$ of length at most $p$, with an end-point in $X$. (Note that our notion of locality is very similar to the classical notion of local algorithm on graphs used in distributed computing, where "the output of a node in a local algorithm is a function of the input available within a constant-radius neighbourhood of the node"~\cite{Suomela13}.)
In that particular setting, Eq. \ref{eq:expanded_expected} can be written as:
\begin{align}\label{eq:expanded_expected_local}
    E_f = \sum_i \bra{\psi^{\Omega(i)}}N^{(i)}\ket{\psi^{\Omega(i)}} +\sum_{\langle i,j \rangle} \bra{\psi^{\Omega(i,j)}} M^{(i,j)}\ket{\psi^{\Omega(i,j)}}
\end{align}
where $\ket{\psi^{\Omega(i)}}$ (resp. $\ket{\psi^{\Omega(i,j)}}$) is the quantum state obtained by running the algorithm on the neighborhood at distance $p$ around vertex $i$ (resp. edge $\langle i,j \rangle$). 

This locality condition for QA is of course too optimistic at this stage. We shall see in the next subsection that, thanks to Lieb-Robinson type inequalities, we can bound the difference between the actual values of energies on nodes and edges on the whole graph, and those obtained by only considering bounded radius balls around a node or an edge, hence obtaining a result very similar to Equation~\ref{eq:expanded_expected_local}.
%\begin{definition}[p-Local Algorithm]
%We call p-local algorithm an algorithm such that the final energy value of a local observable only depends on a neighborhood at distance $p$ around the support of the observable. 
%\end{definition}
%\begin{definition}[$\mathcal{G}_n$]
%Lets $n$ be a natural number. We note $\mathcal{G}_n$ the set of graphs with $n$ nodes.
%\end{definition}
%\begin{definition}[$\mathcal{N}_p$]
%Let $G$ be a graph and $X$ a subset of its nodes. We note $\mathcal{N}_p$(X) the subgraph $\Omega$ of $G$ containing the graph induced by $X$ and all vertices and edges situated on a path of $G$ of length at most $p$, with an end-point in $X$.
%\end{definition}
%\begin{definition}[$u_X$]
%
%\end{definition}
%\begin{definition}[p-local algo]
%$\forall G,G' \in \mathcal{G}_n, \forall (X,X') \subseteq V(G) \times V(G'), \mathcal{N}_p(X) = \mathcal{N}_p(X') \Rightarrow \forall x \in \{0,1\}^n, \mathbb{E}(u_X(x))=\mathbb{E}(u_{X'}(x))$
%\end{definition}

\begin{figure} [ht]
  \centering
  $
  \begin{array}{cccc}
  \begin{tikzpicture}
  \node[circle, draw, inner sep=1pt](v1) at (0,0){i};
  \node[circle, draw, inner sep=1pt](v2) at (0,-1){*};
  \node[circle, draw, inner sep=1pt](v4) at (1,1){*};
  \node[circle, draw, inner sep=1pt](v3) at (-1,1){*};

  \draw (v1) -- (v2); \draw (v1) -- (v4);
  \draw (v1)-- (v3); 
  \end{tikzpicture} &
  \begin{tikzpicture}
  \node[circle, draw, inner sep=1pt](v1) at (-1,0){i};
  \node[circle, draw, inner sep=1pt](v2) at (0,1){*};
  \node[circle, draw, inner sep=1pt](v4) at (1,0){j};
  \node[circle, draw, inner sep=1pt](v3) at (0,-1){*};

  \draw (v1) -- (v2) -- (v4) -- (v3) -- (v1) --(v4); 
  \end{tikzpicture} &  
  \begin{tikzpicture}
  \node[circle, draw, inner sep=1pt](v1) at (-1,0){i};
  \node[circle, draw, inner sep=1pt](v2) at (0,1){*};
  \node[circle, draw, inner sep=1pt](v4) at (1,0){j};
  \node[circle, draw, inner sep=1pt](v5) at (1,-1){*};
  \node[circle, draw, inner sep=1pt](v3) at (-1,-1){*};

  \draw (v4) -- (v1) -- (v2) -- (v4) -- (v5);
  \draw (v1) -- (v3);
  \end{tikzpicture} &
  \begin{tikzpicture}
  \node[circle, draw, inner sep=1pt](v1) at (-1,0){i};
  \node[circle, draw, inner sep=1pt](v2) at (-1,1){*};
  \node[circle, draw, inner sep=1pt](v4) at (1,0){j};
  \node[circle, draw, inner sep=1pt](v5) at (1,-1){*};
  \node[circle, draw, inner sep=1pt](v3) at (-1,-1){*};
  \node[circle, draw, inner sep=1pt](v6) at (1,1){*};

  \draw (v4) -- (v1) -- (v2);
  \draw (v6)-- (v4) -- (v5);
  \draw (v1) -- (v3);
  \end{tikzpicture} \\
  \Omega_0 & \Omega_1   &  \Omega_2 & \Omega_3
  \end{array}
  $
  \caption{Subgraphs of 3-regulars for 1-qubit and 2-qubits neighborhood at distance 1.}
  \label{subgraphs}
\end{figure}
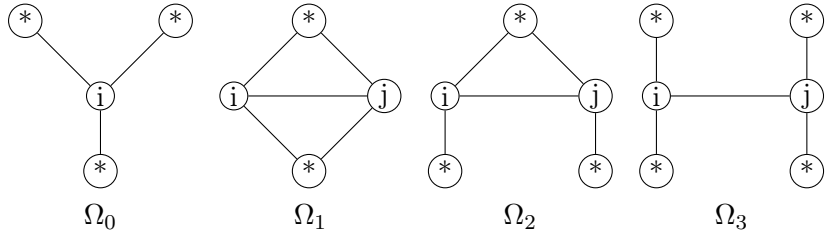

%\subparagraph{Restriction to regular graphs.}
%\marginpar{\ioan{Rappeler que c'est une idée de Farhi et al.?}}
Regular graphs have the particularity of having very few distinct small subgraphs. For instance, the neighborhood at distance $1$ of a vertex is always a star graph. As for edges, their neighborhoods at distance $1$ also form a limited number of configurations, and the number of triangles in such neighborhoods will play a major role. For example, on 3-regular graphs, Fig. \ref{subgraphs} shows the star subgraph $\Omega_0$ at distance one from node $i$ and the three different subgraphs $\Omega_1, \Omega_2$ and $\Omega_3$, at distance one from an edge $\langle i,j \rangle$.

%\subparagraph{Application to 3-regular graphs.} 
Coming back to Eq. \ref{eq:expanded_expected_local}, all the terms in the first sum will be identical since all vertices have the same distance $1$ neighborhood $\Omega_0$. The terms of the second sum can be regrouped into $3$ terms, corresponding to edges having $\Omega_1$, $\Omega_2$, and $\Omega_3$ for distance $1$ neighborhood.
Thus, the expression of $E_f$ becomes:
\begin{align} 
\label{perfect_ef}
     E_f = nE_0 + g_1E_1 + g_2E_2 + g_3E_3
\end{align}
where $g_k$ is the number of edges having $\Omega_k$ for distance $1$ neighborhood, and $E_0 = \langle \psi^{\Omega_0} | N^{(i)} | \psi^{\Omega_0} \rangle$ because $\forall i, \Omega(i) = \Omega_0$, $E_k = \langle \psi^{\Omega_k} | M^{(i,j)} | \psi^{\Omega_k} \rangle$ because $\forall \langle i,j \rangle, \Omega(i,j)$ is one of the $\Omega_k$. The indices $i$ and $i,j$ of the observables refer to the labels on the subgraphs in Fig. \ref{subgraphs}.
We assume that, for configuration $\Omega_1$, the two nodes marked with a $*$ are non-adjacent (if this happens then the four vertices form a connected component of the input graph, which is not critical for our analysis). Lets note $n_k$ the number of configurations $\Omega_k$. Hence there are $n_1=g_1$ edges in configuration 1 and $4n_1 + 3n_2=g_2$ in configuration 2 because there are four sides of $\Omega_1$ and three edges of the triangle in $\Omega_2$, each of them being in a configuration of type $\Omega_2$. The number of edges in configuration $\Omega_3$ corresponds to the remaining ones, i.e., $3n/2 - 5n_1-3n_2=g_3$. Therefore, the final energy can be written as 
\begin{align}\label{eq:reduced_expected}
E_f = n E_0 + n_1 E_1 + (4n_1+3n_2)E_2 +(3n/2 -5n_1-3n_2)E_3
\end{align}
 where $E_k = \langle \psi^{\Omega_k}|O_X|\psi^{\Omega_k}\rangle$ and $O_X$ refers to the appropriate observable $O$ on $X$. %A classical optimization on $n_1/n$ and $n_2/n$ gives a lower bound on the expectation value that this algorithm can reach in worst case scenario.

This construction is the one used to analyze the performances of the QAOA for MaxCut on $3$-regular graphs in \cite{farhi2014quantum}. The authors used Eq. \ref{eq:reduced_expected} together with an upper bound on the maximum cut in $3$-regular graphs in order to derive a worst case lower bound of $0.6924$ on the ratio achieved by a depth $1$ QAOA.
 In our setting of QA, we cannot directly use such construction since the locality condition that allows us to write the final expected cost as in Eq. \ref{eq:expanded_expected_local} does not hold.
To allow the same combinatorial arguments, we thus need to recover some form of locality in the QA framework.

\subsection{Lieb-Robinson like bound}
\label{ssec:lrb}

The main tool we use in this work to regain some relaxed notion of locality in QA is a Lieb-Robinson like bound. Many different versions exist but they all convey the idea of a bounded speed of information in a quantum system. This bounded speed of information entails that, after evolving our system for a short amount of time, a local observation cannot depend too strongly on features lying far from the observed subsystem.
%In our setting, it means that the support of a local observable evolution, like $N_i$ or $M_{ij}$, does not spread too fast, i.e., 
In other words, considering an observable $O_X$ localized on subsystem $X$, the quantity $\langle \psi^G| O_X| \psi^G\rangle$ will be close to $\langle \psi^\Omega| O_X| \psi^\Omega\rangle$ for $\Omega$ a neighborhood of $X$. 
More formally, we want to bound the following quantity:
%, so in order to bound the quantity $|\langle \psi^G |O_X| \psi^G \rangle - \langle \psi^\Omega |O_X| \psi^\Omega \rangle|$, \simon{J'aime pas la phrase suivante.} The goal of this result is to bound $|\langle \psi^G |O_X| \psi^G \rangle - \langle \psi^\Omega |O_X| \psi^\Omega \rangle|$. The first step is to neglect the initial state of norm 1 like this: 
\begin{align*}
    |\langle \psi^G |O_X| \psi^G \rangle - \langle \psi^\Omega |O_X| \psi^\Omega \rangle| 
    &= |\langle \psi_0 |(U_{0,t}^G)^\dagger O_X U_{0,t}^G |\psi_0 \rangle - \langle \psi_0| (U_{0,t}^\Omega)^\dagger O_X U_{0,t}^\Omega |\psi_0 \rangle| \\
    &= |\langle \psi_0 | \left [ (U_{0,t}^G)^\dagger O_X U_{0,t}^G - (U_{0,t}^\Omega)^\dagger O_X U_{0,t}^\Omega \right ]|\psi_0 \rangle|
\end{align*}
Since state-dependent LR bounds does not seem to exist to the best of our knowledge, we first start by neglecting the initial state $\ket{\psi_0}$ of unit norm, and write:
\begin{align*}
    |\langle \psi^G |O_X| \psi^G \rangle - \langle \psi^\Omega |O_X| \psi^\Omega \rangle| \leq \left \lVert (U_{0,t}^G)^\dagger O_X U_{0,t}^G -(U_{0,t}^\Omega)^ \dagger O_X U_{0,t}^\Omega \right \rVert
\end{align*}
where $\left\lVert . \right\rVert$ denotes the operator norm.
The two terms in the norm represent the time evolution of the observable $O_X$ over the whole graph $G$ and over the subgraph $\Omega$. Despite explicit expression of Lieb-Robinson bounds exist like in \cite{hastings2019classical}, they are too loose for our applications on approximation ratios. A numerically tractable expression (not necessarily in a closed form) is enough. 
Although the expression used here is not a LR bound in the strict sense of the concept, it bounds the same quantity, and for simplicity we refer to it as an LR bound.
We will use the following result, adapted from~\cite{Tran_2019} (see proof in Appendix \ref{sec:proof}, inspired from Appendix A of~\cite{Tran_2019}). 

\begin{proposition}[\cite{Tran_2019}]
\label{LR-bound}
Let $\Omega$ be a subgraph of $G$, $H_\Omega$ be the terms of the total Hamiltonian $H_G$ supported on the subgraph, and $O_X$ an observable supported on $X$ included in $\Omega$. The total Hamiltonian $H_G$ is a linear interpolation between $H_0$ and $H_1$ where only $H_1$ has interactions terms. For an evolution during $T$ and $t \in [0,T]$, we have that:
\begin{align}
    \left \lVert (U_{0,t}^G)^\dagger O_X U_{0,t}^G -(U_{0,t}^\Omega)^ \dagger O_X U_{0,t}^\Omega \right \rVert \leq \int_0^t ds \frac{s}{T} \left \lVert [(U^\Omega_{s,t})^\dagger O_X U^\Omega_{s,t}, H_{1,\partial \Omega}]\right \rVert := LR^\Omega_{O_X}(t)\label{eq:LR-bound-eq}
\end{align}
where $H_{1,\partial \Omega}$ is the final Hamiltonian reduced to the border of $ \Omega$. In practice, only the interaction terms $M^{(i,j)}$ are left. Notation $[.,.]$ corresponds to the commutator operation, i.e. for any operators $A$ and $B$, $[A,B]=AB-BA$.
\end{proposition}

\begin{figure}[ht]
    \centering
    \includegraphics[scale=0.5]{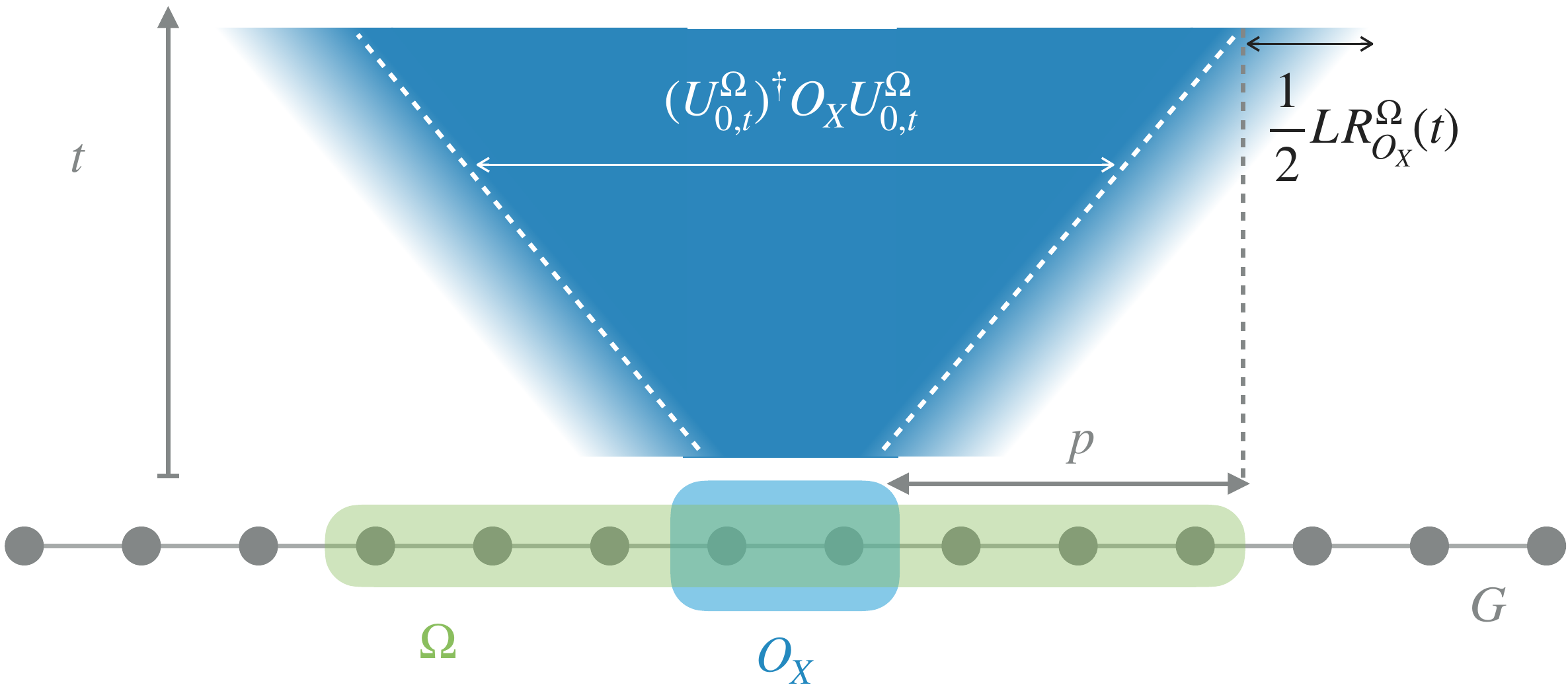}
    \caption{Lieb-Robinson bound illustration on a path graph. The amount of LR bound is in practice quasi null at the beginning and diverges exponentially.}
    \label{fig:LR}
\end{figure}

On the left-hand side of the inequality, we have the norm of the difference between the time evolution of $O_X$ over a graph $G$ with the time evolution of the same observable over a subsystem $\Omega$. This value corresponds to the largest noticeable difference in energy one could possibly measure between the evolution on the full graph or restricted to $\Omega$.
%As seen in the previous subsection, on regular graph, such $\Omega$s with distance one around a node or an edge are the same to any 3-regular graph. Therefore, on the right-hand side, we have a bound that does not depend on the graph $G$ anymore. 
Interestingly, the right-hand side of this inequality (i.e. the LR type bound) does not depend on the whole graph $G$. Potentially, only $H_{1,\partial \Omega}$ could depend on the graph but when working with regular graphs, there are only a few possible choices for this term which depend on $\Omega$. This bound only depends on the local shape $\Omega$ of the system around $X$.
Thus, looking at the value of a local observable of an edge or a node on the whole graph or only locally on a subgraph is the same up to a small error given by this bound. 
%We clearly see that there is a relation between the amount of time $t$ we let the system evolve and the distance $p$ of the neighborhood $\Omega$. 
This LR bound will quickly diverge when letting the evolution run for a large amount of time. In particular, this entails that this result can only be used to analyze the performances of very short time-continuous evolutions.
%Increasing the size of $\Omega$ will produce a slower LR bound, thus refining the analysis while increasing the combinatorial and numerical cost of the analysis.
%, while, in practice, one would tend to let a quantum annealing run for as long as possible in order to maximize the quality of the produced solutions.
Moreover, the time dependence of the bound is such that considering a larger subsystem $\Omega$ will yield a smaller bound for a fixed $T$. In other words, if one needs to study some evolution during a longer time (e.g. to get closer to the exact solution), one might need to increase the size of $\Omega$.
%Thus, there is a trade-off between letting the evolution for a large amount of time to get closer to the exact value and not too large to be able to simulate the evolution on small subgraphs. 
Fig. \ref{fig:LR} illustrates the time evolution of an observable $O_X$ on a path graph. The light cone of the information spreading is colored in blue. 
In this work, we choose to work with balls of radius $p=1$ around each vertex and edge, hence the subgraphs of Fig. \ref{subgraphs}, so we expect to have a very short running time $T$.

Now, with this result, we can lower bound the value of $\langle \psi^G |O_X| \psi^G \rangle$ that appears in the expression of the final energy $E_f$. So we have a lower bound that is graph-independent on the expected value of the quantum state output by QA. Choosing a tight upper bound on the optimal value will give a global lower bound of the approximation ratio. Indeed, we have that $\langle \psi^G |O_X| \psi^G \rangle \geq \langle \psi^\Omega |O_X| \psi^\Omega \rangle - LR^\Omega_{O_X}$, so with $E_k^* = \langle \psi^{\Omega_k} |O_X| \psi^{\Omega_k} \rangle - LR^{\Omega_k}_{O_X}$, the approximation ratio can be computed via: 

\begin{align}\label{eq:rho3}
\rho(T) = \frac{E_f}{C_{\hbox{opt}}} \geq  \frac{n E_0^* + n_1 E_1^* + (4n_1+3n_2)E_2^* +(3n/2 -5n_1-3n_2)E_3^* }{C_{\hbox{opt}}}
\end{align}

In this section, we have shown that restricting the combinatorial problem to regular graphs helps to study locally the problem because the number of small subgraphs is finite. Then, thanks to the Lieb-Robinson bound, we can lower bound the expected value of a local observable on the whole graph which is needed to compute the final expected value of QA's quantum state. In practice, we shall use Proposition~\ref{LR-bound} to calculate the LR type bound numerically.
In the next section, we will apply this construction to two combinatorial problems on graphs: MaxCut and Maximum Independent Set (MIS). 

\section{Application to MaxCut and MIS}
\label{sec:app}
In this section, we apply the method suggested to derive approximation ratios for MaxCut and MIS on 3-regular graphs using QA metaheuristic. In both cases, we need first to explicit the cost function, then to encode it in the Hamiltonian $H_1$. We then proceed by numerically evaluating the energies $E_k^*(T)$ on the four subgraphs $\Omega_k$ described in Fig. \ref{subgraphs}, and by optimizing the ratio with respect to the annealing time $T$. 

For $k \in \{0,1,2,3\}$, we compute the value of $E_k^*(T)$ for $T \in [0,5]$. To do this, we need the value of $E_k(T)$ and the Lieb-Robinson like bound $LR_{O_X}^{\Omega_k}(T)$ for appropriate $O_X$.
$E_k(T)$ can be computed directly from the quantum state $\ket{\psi^{\Omega_k}(T)}$. This state is computed by numerically solving  the Schr\"odinger equation (\ref{eq:shrod}) with a relative tolerance of $10^{-9}$ on $\Omega_k$. These computations were performed on an Atos QLM ({Quantum Learning Machine}) using boost. To compute the LR bound, we have to compute the unitary evolution operator $U_{s,T}^{\Omega_k}$ for all $s \in [0,T]$ and compute the integral on the right hand side of Eq. \ref{eq:LR-bound-eq}. We chose a time step of $10^{-3}$ to compute the integral. We compute the unitary operator for each time step by solving the Schr\"odinger equation for evolution operator (Eq.~\ref{eq:shrod2}). These computations were carried using the Differential Equations Julia library.

Let us detail the interaction terms $H_{1,\partial \Omega}$. This is a sum over all interactions $M^{(i,j)}$ for edges $\langle i,j\rangle$ which are not in $\Omega$ but have at least one endpoint in $\Omega$. Therefore, the larger the sum is, the larger the LR bound will be. Considering interactions between two nodes of $\Omega$ reduces the bound because we work with bounded degree graphs. So the worst case considers only interactions between $\Omega$ and $\bar{\Omega}$, i.e. interactions that leave the subgraph. To include the $\Omega_k$'s in cubic graphs, there are 6 similar terms with $\Omega_0$, 2 same terms when considering $\Omega_1$, 5 terms including 4 identical ones with $\Omega_2$ and 8 similar terms with $\Omega_3$.

\subsection{MaxCut}

The cost function $C$ we want to maximize counts the size of a cut, i.e. the number of edges cut by a solution $x=x_1x_2...x_n$, where $x_i \in \{0,1 \}$ (nodes $i$ such that $x_i = 0$ are considered as on the left-hand side of the cut, while nodes $j$ with $x_j = 1$ are on the right-hand side). So we have $C(x)=\sum_{\langle i,j\rangle } x_i \oplus x_j$, a sum of xor operations between the extremities of each edge.  By a change of basis from $x_i \in \{0,1\}$ to $z_i \in \{1,-1\}$, the cost function becomes $C(z) = \sum_{\langle i,j\rangle} \frac{1}{2}(1-z_iz_j)$. In quantum computing, there is a natural operator that does this basis change called $\sigma_z$. Therefore, the natural expression for the final Hamiltonian that QA will minimize is $H_1 = -\sum_{\langle i,j\rangle} \frac{1}{2}(1- \sigma_z^{(i)} \sigma_z^{(j)})$. Thus, for MaxCut, we have $N^{(i)}=0$ and $M^{(i,j)}=\frac{1}{2}(1-\sigma_z^{(i)} \sigma_z^{(j)})$. 
For the optimal cut, we use the same reasoning in \cite{farhi2014quantum}: the size of the maximum cut is at most $3n/2$ and each configuration of type $\Omega_1$ and $\Omega_2$ (see Figure~\ref{subgraphs}) has at least one edge that is not cut (they each contain a triangle that is disjoint from all other triangles). Hence, we have that $C_{\hbox{opt}} \leq 3n/2 - n_1 - n_2$. By normalizing with $n_1 \leftarrow n_1/n$ and $n_2 \leftarrow n_2/n$, we end up with a global approximation ratio $\rho_{mc}$ for problem MaxCut that verifies:

\begin{align}\label{eq:rmc2}
    \rho_{mc} \geq \min_{\begin{array}{c}
         n_1,n_2 \text{ s.t.} \\
         4n_1+3n_2 \leq 1
    \end{array} } \max_T \frac{n_1 E_1^*(T) + (4n_1+3n_2)E_2^*(T) +(3/2 -5n_1-3n_2)E_3^*(T) }{3/2 - n_1-n_2}
\end{align}
By numerically evaluating the $E_k^*(T)$ and optimizing over $T, n_1, n_2$, we find that the approximation ratio reaches 0.5933 at $T_{mc}=1.62$ and $n_1=n_2=0$. The numerical values for $E_k^*$ are presented in Table \ref{tab:mcval}, we can see that $ \rho_{mc}\geq E_3^*$. This means that QA struggles the most with triangle-free graphs, as it also happens with QAOA~\cite{farhi2014quantum}. A refined numerical analysis is presented in Appendix \ref{app:hints} where we point out more intuitions about QA on MaxCut.

\begin{table}[ht]
    \centering
    \begin{tabular}{c|c|c|c}
        k & 1 & 2 & 3 \\
        \hline
        $E_k$ & 0.5951 & 0.6152 & 0.6350 \\
        $LR_{O_X}^{\Omega_k}$ & 0.0203 & 0.0310 & 0.0417 \\
        $E_k^*$ & 0.5748 & 0.5842 & 0.5933
    \end{tabular}
    \caption{Numerical values to get $E_k^*(T_{mc})$}
    \label{tab:mcval}
\end{table}

%\subparagraph{Intuition} 
%Let us try to explain the result. Imagine the observable $O_X$ as some information localized on $X$, and we look at how it spreads in the graph during a QA evolution. The more spread out the graph is, the farther (in the graph) the information must travel. By "spread out", we mean that there is no small "cycle", so no triangle at least. The same conclusion results from the LR-bound of Theorem~\ref{LR-bound}. Indeed, the bound depends on the number of interactions at the subgraph $\Omega$'s border. Since our graphs are regular, this number is maximized when $\Omega$ induces a subtree in $G$. We will detail and develop this idea in Section~\ref{sec:exp}.

\subsection{Maximum Independent Set}
For MIS problem, we use the same cost function as in~\cite{farhi2020quantum}. For a vector $x=x_1x_2...x_n$, with $x_i \in \{0,1 \}$, the cost function tries to maximize the Hamming weight and minimize the number of edges in the subset $x$, i.e., the subset of nodes $i$ such that $x_i = 1$. We write $ C(x) = \sum_{i } x_i - \sum_{\langle i,j\rangle} x_ix_j$. 
Even though this function only weakly encodes the independent set constraint, one can easily transform any bit-string $x$ of cost $C(x)$ into a stable of size at least $C(x)$. This can be done considering each edge that violates the stable condition (i.e. each edge $\langle i, j \rangle$ such that $x_i = x_j = 1$), picking at random one of its extremities, and removing it from the solution. This effectively removes a vertex from $x$ while "fixing" at least an edge, thus creating a new solution with an increased cost. Eventually, the maximum of $C(x)$ corresponds to a maximum independent set.
%An optimal solution, maximizing $C$, is not necessarily an independent set but can easily be transformed into an independent of the same size size $C(x)$. From $x$ to recover an IS, we remove randomly one of the two nodes that form an edge in $x$. 

In order to implement $C$ into a Hamiltonian we proceed to the same change of basis as for MaxCut: $\{0,1\}$ is transformed into $\{1,-1\}$ and $C$ can be written as $C(z)=\sum_{i } \frac{1-z_i}{2} - \sum_{\langle i,j \rangle } \frac{1-z_i}{2}\frac{1-z_j}{2}$. Thus, the final Hamiltonian for MIS that QA will minimize is: $H_1 = -\sum_i \frac{1}{2}(1-\sigma_z^{(i)}) + \sum_{\langle i,j \rangle} \frac{1}{4}(1-\sigma_z^{(i)}) (1-\sigma_z^{(j)})$. Unlike MaxCut, we see here that we have a component on the nodes and a component on the edges; $N^{(i)}=\frac{1}{2}(1-\sigma_z^{(i)})$ and $M^{(i,j)}=-\frac{1}{4}(1-\sigma_z^{(i)})(1-\sigma_z^{(j)})$. In order to upper bound the size of an IS in regular graph, we use the following claim.

\begin{claim}
For a $d$-regular graph, the size of the MIS is upper bounded by $n/2$ and this value is reached by bipartite graphs.
\end{claim}
The proof is straightforward once we see that the total number of edges in a $d$-regular graph is $\frac{dn}{2}$, and any independent set $I$ has $d|I|$ edges incident to it.

To compute the approximation ratio of MIS on 3-regular graphs, we will use this upper bound on $C_{\hbox{opt}}$. By the same normalization with $n_1 \leftarrow n_1/n$ and $n_2 \leftarrow n_2/n$, we end up with the following bound on the approximation ratio:

\begin{align}
    \rho_{mis}\geq\min_{\begin{array}{c}
         n_1,n_2 \text{ s.t.} \\
         4n_1+3n_2 \leq 1
    \end{array} } \max_T \left ( 2(E_0^*+n_1 E_1^* + (4n_1+3n_2)E_2^* +(\frac{3}{2} -5n_1-3n_2)E_3^* ) \right )
\end{align}
By numerically evaluating the $E_k^*(T)$ and optimizing over $T, n_1, n_2$, we find that the approximation ratio reaches 0.3171 at $T_{mis}=1.32$ for $n_1=n_2=0$. The numerical values for $E_k^*$ are presented in Table \ref{tab:misval}, and we have that $\rho_{mis}\geq 2E_0^*(T_{mis})+3E_3^*(T_{mis})$.

%\ioan{Donner toutes les valeurs $E_k^*$. On ne coupera pas à un commentaire genre "This approximation ratio is no better than the $0.5$ factor achieved by the greedy algorithm for cubic graphs, nonetheless our purpose  is to illustrate the generality of the approach based on Lieb-Robinson bounds, which can be used for estimating the energy both on vertices and on edges.} \arthur{est ce qu'on sait comment SA se comporte pour approcher MIS? comme le "random guess" n'est pas bien défini, QA "marche bien" dans le sens où il fait mieux que 0.25.}

\begin{table}[ht]
    \centering
    \begin{tabular}{c|c|c|c|c}
        k & 0 & 1 & 2 & 3 \\
        \hline
        $E_k$ & 0.4653 & -0.1939 & -0.1920 & -0.1901 \\
        $LR_{O_X}^{\Omega_k}$ & 0.0072 & 0.0034 & 0.0065 & 0.0096 \\
        $E_k^*$ & 0.4581 & -0.1972 & -0.1985 & -0.1997
    \end{tabular}
    \caption{Numerical values to get $E_k^*(T_{mis})$}
    \label{tab:misval}
\end{table}

Our purpose is to illustrate the generality of the approach based on Lieb-Robinson bounds, which can be used for estimating the energy both on vertices and on edges. We reach with a constant-time local algorithm a ratio that is better than random guess. Here, the random guess is among all bitstrings $x \in \{0,1\}^n$ so the expectation of $C(x)$ (thus the expected size of MIS) is of $n/8$, which gives an approximation ratio of 0.25. \\

%\noindent In this section, we show how to apply the general method of using an LR-bound with QA to compute an approximation ratio with a constant time algorithm local at distance 1. 
%Even though, the goal of this paper is to show a new analytical method to compute approximation ratio, we believe that it can actually reach far better numerical values, in the next section we give our intuition over MaxCut.

\section{Toward a better bound for MaxCut?}
\label{sec:exp}

In the previous section, we have shown how to apply a general method using an LR bound with QA to compute an approximation ratio with a constant time, local algorithm (at distance 1), for cubic graphs.

%intuition : on pense que QA fait mieux et on essaie d'argumenter pq (mieux structurer) puis pq les bornes sont relachées. 

We believe that the algorithm behaves significantly better than the approximation bounds that we were able to prove. In Subsection~\ref{ssec:estimations}, we give a theoretical algorithm to find a tighter bound on the approximation ratio and argue that the actual approximation ratio for MaxCut in cubic graphs with a distance 1 analysis might be of $0.6963$, while we only guarantee $0.59$. Then, in Subsection~\ref{ssec:tightness}, we discuss the (non) tightness of our LR type bound used in the analysis, explaining this gap, and suggesting some directions for improvements.

\subsection{Hints for better approximation ratios}\label{ssec:estimations}

To derive a better bound with a distance 1 analysis, i.e. still using Eq. (\ref{eq:reduced_expected}), ideally we should use energies $E_k^{\mathcal{G}}$, corresponding to the minimum energy of the edge $X$ in $\Omega_k$ among all possible cubic graphs containing $\Omega_k$:
$$E_k^\mathcal{G}(T) = \min_{G\text{ s.t. } X \in \Omega_k \subseteq G} \bra{\psi^G(T)}O_X \ket{\psi^G(T)}. $$ 

Instead we used values $E_k^*$ that are lower bounds for $E_k^{\mathcal{G}}$, obtained through our LR-type bound of Proposition~\ref{LR-bound}. In order to improve the approximation ratio, we
need to access to a more accurate estimation of values $E_k^{\mathcal{G}}$.
Values $E_k^{\mathcal{G}}$ are obviously intractable in practice because there are infinitely many cubic graphs $G$ that can complete $\Omega_k$. 

However, we can make a different use of LR bounds to reduce the search space to the finite space of all balls $\mathcal{B}$ that can be completed in a 3-regular graph with finite radius $r$ (from the edge $X$). Here we employ \cite[Lemma 5]{Haah_2021} stating that there are some constants $v_{LR} > 0$ (called the Lieb-Robinson velocity) and $\mu > 0$ such that for any $r \geq v_{LR}t$, there exists a ball $B_r$ centered on $X$ with radius $r$ such that:
\begin{equation}
    \left \lVert (U_{0,T}^G)^\dagger O_X U_{0,T}^G -(U_{0,T}^{B_r})^ \dagger O_X U_{0,T}^{B_r} \right \rVert \leq \mathcal{O}(e^{-\mu r})
\end{equation}
Thus, by only considering balls of radius $r$, we get can get an estimate of the true value of $E^*_k(T)$ up to $\mathcal{O}(e^{-\mu r})$.

%By looking only at the balls of radius $r$, at worst we induce an error of order $\mathcal{O}(e^{-\mu r}$
For any constant time $T$, in bounded degree graphs, this statement allows us to fix an arbitrarily small error $\varepsilon >0$ which fixes a radius $r(\varepsilon, T)$. From there, we can approximate the energy of a vertex or an edge at time $T$ of a QA process by only studying a neighborhood at constant distance around the vertex or the edge. The corollary is an application of Theorem 1 in \cite{moosavian2021limits}, it can also be deduced from Lemma 5 of \cite{Haah_2021}.

\begin{corollary}
\label{cor:lr}
Let $G$ be a bounded degree graph, $O_X$ an observable supported on X and let $B_r$ denote the neighborhood of $X$ at distance at most $r$. For any $\varepsilon > 0$ and short runtime $T$, there exists $r(\varepsilon, T)$ such that :
\begin{align}
    |E_G(T) - E_{B_{r(\varepsilon, T)}}(T)| \leq \varepsilon 
\end{align}
where $E_\Lambda(T)=\bra{\psi^\Lambda(T)}O_X \ket{\psi^\Lambda(T)}$
\end{corollary}
Let us come back to our distance $1$ analysis of MaxCut on cubic graphs.
This corollary entails that one can pick a precision $\varepsilon$ and a time $T$ and obtain a distance $r(\varepsilon, T)$ such that for all balls of distance $r(\varepsilon, T)$ around an edge, any completions of this ball into a full cubic graph will not change the expected value of the energy of this edge by more than $\varepsilon$. This gives us an effective (but not realistic) way to find a graph-independent lower bound on $E_k^{\mathcal{G}}(T)$ up to an additive error of $\varepsilon$:
\begin{enumerate}
    \item consider each type of distance $1$ neighborhood $\Omega_k$, for $k\in \{1,2,3\}$,
    \item for each completion of $\Omega_k$ into a distance $r(\varepsilon, T)$ ball $B_r$, compute $\bra{\psi^B}O_X\ket{\psi^B}$,
    \item pick the lowest value $E_k^\mathcal{B}(T) = \min_{B\text{ s.t. } X \in \Omega_k \subseteq B} \bra{\psi^B}O_X\ket{\psi^B}, $ and output $E_k^\mathcal{B} - \varepsilon$ as lower bound for $E_k^\mathcal{G}(T)$.
%    \item use Eq. \ref{eq:rmc2} to produce a lower bound of $\rho_{mc}$ using $E_k^* = E_k^{min} - \varepsilon$.
\end{enumerate}

%\begin{proposition}
%Let us fix $\varepsilon$ and $T$ and consider the distance $r(\varepsilon, T)$ as in corollary \ref{cor:lr}.
%For all cubic graph $G$ and for all edge $\langle i,j\rangle$ in $G$ with distance $1$ neighborhood $\Omega_k$, we have that:
%\begin{align}\label{eq:prop_lower_bound}
%    \bra{\psi^G} M_{i, j} \ket{\psi^G} \geq  E_k^{min} - \varepsilon
%\end{align}

%where $E_k^{min}$ is the value produce by the algorithm above. Consequently, this algorithm produces a valid lower bound of $\rho_{mc}$ for a given annealing time $T$.
%\end{proposition}
%\begin{proof}
%For each graph $G$ and each edge $\langle i, j \rangle$ in $G$, consider the ball $\mathcal{B}$ of distance $r(\varepsilon, T)$ around $\langle i,j\rangle$. We have that $\bra{\psi^\mathcal{B}} O_X \ket{\psi^\mathcal{B}} \geq E_k^{min}$ by definition of $E_k^{min}$ and we have that $|\bra{\psi^G} O_X \ket{\psi^G} - \bra{\psi^\mathcal{B}} O_X \ket{\psi^\mathcal{B}}| \leq \varepsilon$ by definition of $r(\varepsilon, T)$. A simple triangular inequality gives us Eq. \ref{eq:prop_lower_bound}.
%\end{proof}

These estimations $E_k^\mathcal{B} - \varepsilon$ are tighter than the ones in Section \ref{sec:app}, basically thanks to the use of larger balls.
Once equipped with these values for each $k\in\{1, 2, 3\}$, we can use Eq. \ref{eq:rmc2} to compute a tighter lower bound than the one derived in Section \ref{sec:app} for the approximation ratio $\rho_{mc}$.
This algorithm is well defined but not practical because there are exponentially many balls to enumerate w.r.t. the quantity $r(\varepsilon, T)$, which might be quite large even for reasonably small values of $T$ and of $\varepsilon$.

\begin{figure} [ht]
  \centering
  $
  \begin{array}{cccc}

  \begin{tikzpicture}
  \node[circle, draw, inner sep=1pt](v1) at (-0.25,0){1};
  \node[circle, draw, inner sep=1pt](v2) at (0,1){3};
  \node[circle, draw, inner sep=1pt](v4) at (1,0){2};
  \node[circle, draw, inner sep=1pt](v3) at (0,-1){4};
  \node[circle, draw, inner sep=1pt](v5) at (-1,0){5};

  \draw (v1) -- (v2) -- (v4) -- (v3) -- (v1) --(v4); 
  \draw (v2) -- (v5) -- (v3);
  \end{tikzpicture} &  
  \begin{tikzpicture}
  \node[circle, draw, inner sep=1pt](v1) at (-1,0){1};
  \node[circle, draw, inner sep=1pt](v2) at (0,1){4};
  \node[circle, draw, inner sep=1pt](v4) at (1,0){2};
  \node[circle, draw, inner sep=1pt](v5) at (1,-1){5};
  \node[circle, draw, inner sep=1pt](v3) at (-1,-1){3};
  \node[circle, draw, inner sep=1pt](v6) at (0,-0.5){6};
  \node[circle, draw, inner sep=1pt](v7) at (0,-1.5){7};

  \draw (v4) -- (v1) -- (v2) -- (v4) -- (v5);
  \draw (v1) -- (v3) -- (v6) -- (v2); 
  \draw (v6) -- (v5) -- (v7) -- (v3);
  \end{tikzpicture} &
  \begin{tikzpicture}
  \node[circle, draw, inner sep=1pt](v1) at (-1,0){1};
  \node[circle, draw, inner sep=1pt](v2) at (-1,1){4};
  \node[circle, draw, inner sep=1pt](v4) at (1,0){2};
  \node[circle, draw, inner sep=1pt](v5) at (1,-1){5};
  \node[circle, draw, inner sep=1pt](v3) at (-1,-1){3};
  \node[circle, draw, inner sep=1pt](v6) at (1,1){6};
  \node[circle, draw, inner sep=1pt](v7) at (0,1.5){7};
  \node[circle, draw, inner sep=1pt](v8) at (0,-1.5){8};
  \node[circle, draw, inner sep=1pt](v9) at (-0.1,-0.25){9};

  \draw (v4) -- (v1) -- (v2) -- (v7) -- (v6) -- (v8) -- (v3) -- (v7);
  \draw (v6)-- (v4) -- (v5) -- (v9) -- (v2);
  \draw (v1) -- (v3); \draw (v8) -- (v5);
  \end{tikzpicture} &
  \includegraphics[scale=0.25]{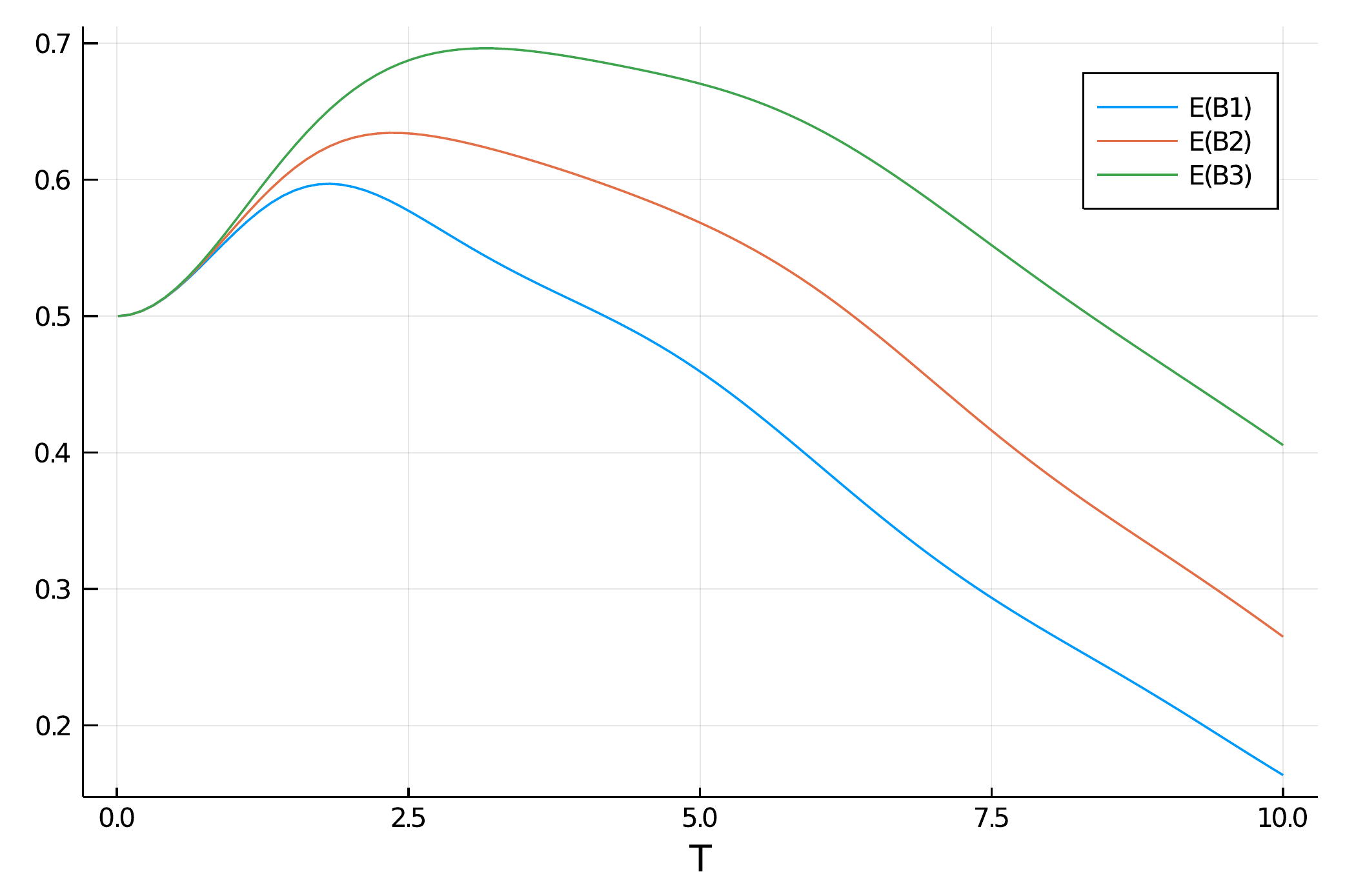} \\
  B_1   &  B_2 & B_3 & \text{Simulations}
  \end{array}
  $
  \caption{Balls $B_k$ of 3-regular graphs around the edge $\langle 1,2 \rangle$ in corresponding configuration $\Omega_k$. On the right, evolution of the edge $\langle 1,2 \rangle$'s energy against $T$, E(B1) in blue, E(B2) in red and E(B3) in green. }
  \label{fig:worstB}
\end{figure}
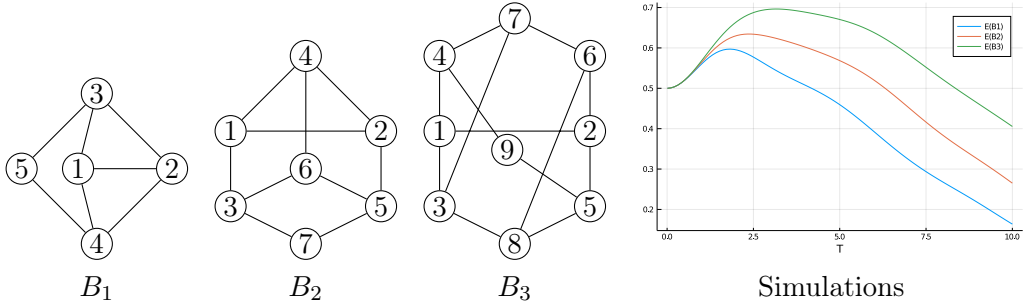

That being said, we still tried to numerically investigate the values we could reach for $E_k^\mathcal{B}$ in the case of MaxCut. 
We wish to obtain the minimum energies $E_k^\mathcal{B}$ for each of the three types of edges, $k \in \{1,2,3\}$.
Intuitively, these minimum energies are reached by edges that are necessarily uncut by the optimal cut. Indeed, these edges see their energies converge to 0 after the initial rise due to locality (experienced by all edges).
%A very simple and illustrative case are the 2-regular graphs where we can clearly see how the locality is expressed in a short amount of time. We show the details in Appendix~\ref{sec:cycle}. 
Thus, our strategy was to construct (small) balls $B_k$ for each configuration $\Omega_k$ such that the energy of the observed edge quickly goes to 0 after its expected initial rise.
%, i.e. where the target edge is uncut in the optimal cut. 
In order to accelerate the convergence speed to 0, we enforced this condition via the adjunction of the smallest possible odd cycles. Indeed, as we can see in Appendix \ref{sec:cycle}, small cycles converge faster to their target. The proposed balls $B_k$ for each configuration $\Omega_k$ are shown in Fig. \ref{fig:worstB}. We computed the energy of the edge $\langle 1,2 \rangle$ in each of these balls for $T \in [0,5]$ and used Eq. \ref{eq:rmc2}, giving us a ratio of 0.6963 at $T^*=3.15$. As mentioned above, this is just a numerical estimate and by no mean a formal derivation. However, unless some very non-monotonous behavior arises when growing the radius of these balls, this is most likely an accurate estimate.
%As mentioned in Subsection~\ref{ssec:estimations}, this is just a numerical estimate, not a formal derivation. However, probably a faithful estimate.

%\begin{remark}
%These suggested balls are isolated from the rest of %the whole graph because only one node is missing one %edge. This entails that the neighborhood at distance %3 is just one edge, so the completion in a cubic %graph will not change the observed edge properties.
%\end{remark}

% \begin{remark*}
% The initial raise can be explained by the fact that an edge is always cut by the optimal one. In these balls, the observed edge always belongs to two odd cycle to force the uncut, after some time (but not too long).
% \end{remark*}

This construction is a truly local (distance 1) analysis of QA, and computing the values of $E_k^\mathcal{B}$ would produce the best possible estimate of QA's performances in this setting. It thus seems reasonable to compare this ratio to one obtained for other truly local algorithms. The other known bounds in this setting are depth-1 QAOA ($\rho = 0.6925$)~\cite{farhi2014quantum} and Hastings' (classical) local algorithm ($\rho=0.6980$)~\cite{hastings2019classical}. If our bound is verified, this would show that a QA in time $T^*=3.15$ outperforms depth-1 QAOA. Additionally, if the bound is tight, Hastings' classical algorithm still beats QA as analyzed in this regime.

\subsection{(Non) Tightness of our LR bound}\label{ssec:tightness}
The analysis of the LR bound tightness or non-tightness depends on the runtime $T$ and the size of $\Omega$, more precisely on its radius $p$. 
%In usual closed forms of the bound like in \cite{hastings2010localityBC}, it increases exponentially fast with $t$ and decreases exponentially fast with $p$ (at different rates). In the application of Section \ref{sec:app}, we restrict ourselves to $p=1$, so we will only look at the tightness in this case, and 
We study the dependency in $T$. We can consider the tightness at $T_{mc}$ where we reach the best ratio using LR bound but we can also wonder why we do not reach something higher for larger $T$ up to $T^*$. \par
At $T_{mc}$, without considering LR bound, i.e. using the final energy of (\ref{perfect_ef}), we reach an approximation ratio of 0.635 still with $n_1=n_2=0$, so it gives a relative error of $6.6 \%$. To understand how to improve our ratio up to $T^*$, we are able to point out which lose have the more impact when deriving the bound. There are two steps where we neglect information. The first one is when we neglect the initial state $\ket{\psi_0}$ at the beginning of section \ref{ssec:lrb} i.e. between $\delta_1(G)=|\langle \psi^{G} |O_X| \psi^{G} \rangle -\langle \psi^{\Omega_3} |O_X| \psi^{\Omega_3} \rangle| $ and $\delta_2(G)=  \lVert (U_{0,t}^G)^\dagger O_X U_{0,t}^G - (U_{0,t}^{\Omega_3})^\dagger O_X U_{0,t}^{\Omega_3}  \rVert$ where here $O_X$ represents the observable $M_{ij}$ of an edge in $G$. The second one is when proving Proposition~\ref{LR-bound} we use the triangular inequality with the operator norm and the integral, i.e., between $\delta_2(G)$ and $LR_{O_X}^{\Omega_3}$.

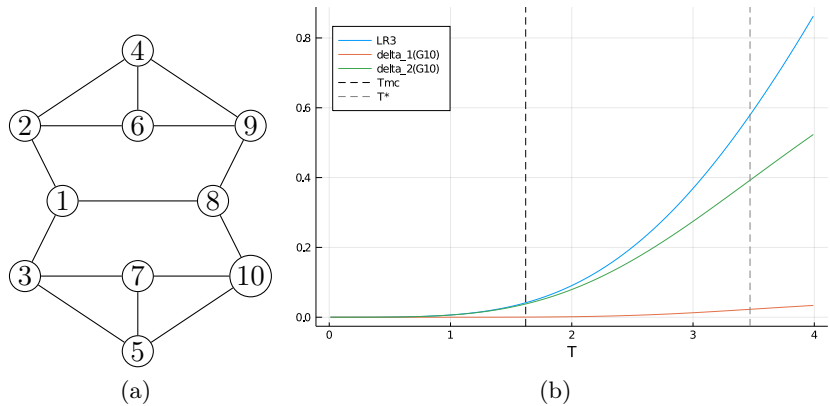
\begin{figure}[ht]%
 \centering
 \subfloat[]{
 \begin{tikzpicture}
        \node[circle, draw, inner sep=1pt](v1) at (-1,0){1};
        \node[circle, draw, inner sep=1pt](v2) at (-1.5,1){2};
        \node[circle, draw, inner sep=1pt](v3) at (-1.5,-1){3};
        \node[circle, draw, inner sep=1pt](v4) at (0,2){4};
        \node[circle, draw, inner sep=1pt](v5) at (0,-2){5};
        \node[circle, draw, inner sep=1pt](v6) at (0,1){6};
        \node[circle, draw, inner sep=1pt](v7) at (0,-1){7};
  
        \node[circle, draw, inner sep=1pt](v8) at (1,0){8};
        \node[circle, draw, inner sep=1pt](v9) at (1.5,1){9};
        \node[circle, draw, inner sep=1pt](v10) at (1.5,-1){10};
        
        \draw (v1) -- (v8) -- (v9) -- (v4) -- (v2) -- (v1) -- (v3) -- (v5) -- (v10) -- (v8) ; 
        \draw (v3) -- (v7) -- (v10); \draw (v6) -- (v4);
        \draw (v2) -- (v6) -- (v9); \draw (v7) -- (v5);
    \end{tikzpicture}\label{fig:5a}}
 \subfloat[]{
 \includegraphics[scale=0.35]{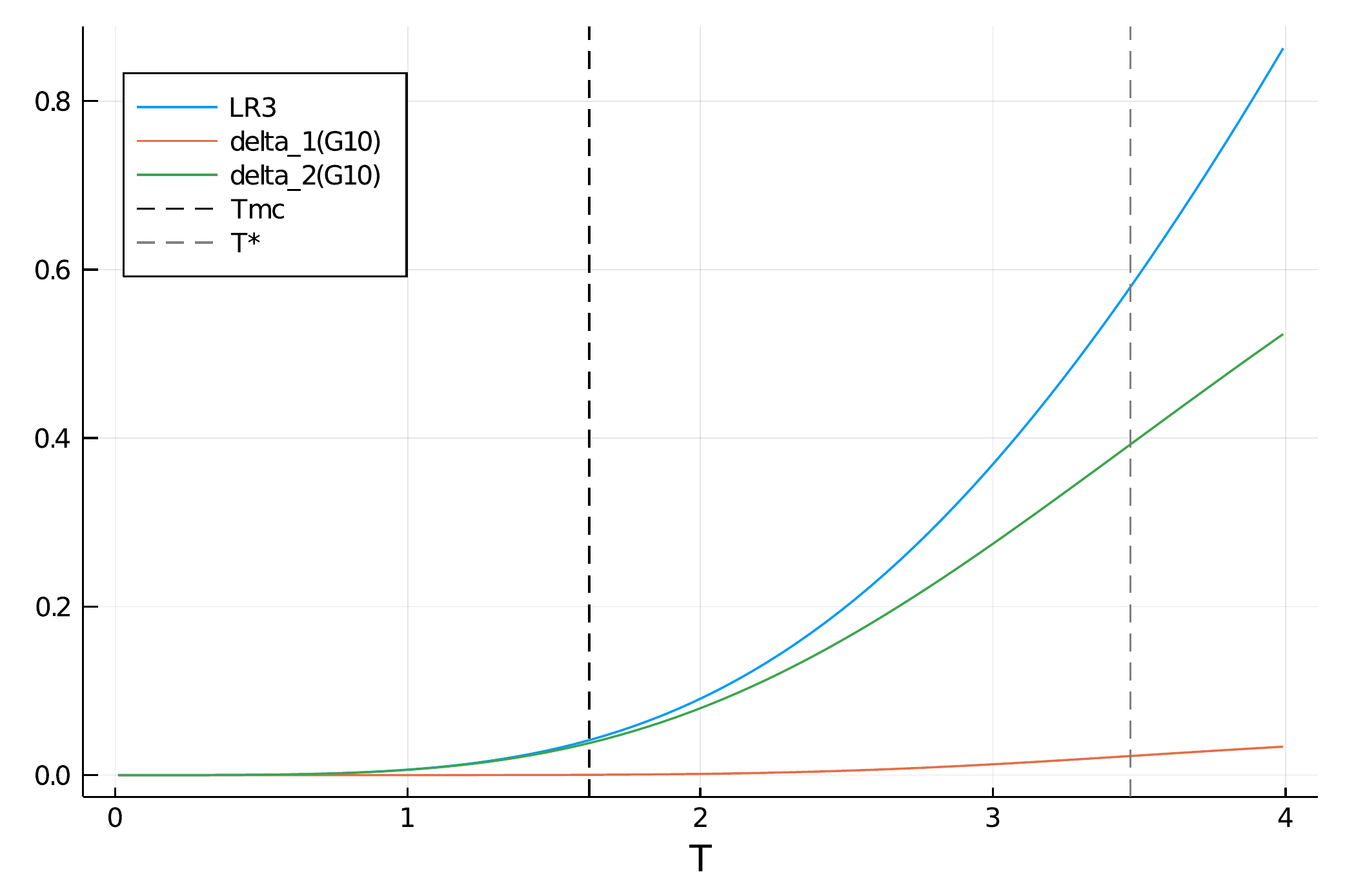}\label{fig:5b}}%
 \caption{On the left (a), $G_{10}$, worst edge configuration for LR bound centered on $\langle 1,8 \rangle$. On the right, $\delta_1$ and $\delta_2$ for $G_{10}$ aside with LR3=$LR_{O_X}^{\Omega_3}$ for $O_X=M_{1,8}$ The dotted lines indicate the time $T_{mc}$ for which we obtain the value of 0.59 and $T^*$.}%
 \label{fig:tight}%
\end{figure}

To illustrate the latter point and the non-tightness of the LR-type bound used, we plotted in Fig. \ref{fig:tight} these three quantities $\delta_1(G) \leq \delta_2(G) \leq LR_{O_X}^{\Omega_3}$ for a non-trivial example.

%We only consider $\Omega_3$ because we know that this is the worst configuration from a QA point of view when solving MaxCut and believe it will still be the case with tighter bounds. We also know from computing the $LR_{O_X}^{\Omega_3}$ bound that the worst case is when there are 8 interactions with the outside of $\Omega_3$. Then, we need a graph such that when computing an edge energy $\langle \psi^{G} |O_X| \psi^{G} \rangle$ it diverges quickly from $E_3=\langle \psi^{\Omega_3} |O_X| \psi^{\Omega_3} \rangle$. Either it goes faster toward 1, or it does not converge to 1 and increases less rapidly. Such graphs are the ones that have more information to decide where to converge. These correspond to small graphs because it understands its structure more quickly. The smaller graphs with 8 interactions are of size 10, indeed, these 8 edges can be produced by only 4 other nodes. The chosen graph is presented on Fig. \ref{fig:5a} where the edge $\langle 1,8 \rangle$ respects the above conditions. Its MaxCut does not cut this specific edge whereas on $\Omega_3$, the middle edge is cut. This is why it creates a larger gap. In Fig. \ref{fig:5b}, we can see the three curves we want to compare, namely $LR_{O_X}^{\Omega_3}$, $\delta_1(G_{10})$ and $\delta_2(G_{10})$. 

The LR bound used to derived the approximation ratio is interesting up to $T_{mc}$, where there is a loss of only $6.6 \%$. But we cannot evolve longer than this runtime because it diverges too quickly. We see that $\delta_2$ also diverges but not $\delta_1$. This means that neglecting the initial state strongly deteriorates the bound.

\section{Conclusion}
\label{sec:discussion}
%\subsection{Conclusion}
The general method described in this article provides tools to derive approximation ratios for quantum annealing on combinatorial problems, with very short running time. The locality in quantum annealing given by the Lieb-Robinson bound allows combinatorial arguments over the regular (and in particular cubic) graphs, of the same flavour as the ones used for QAOA. We emphasize that the theoretical arguments hold for regular graphs of any fixed degree, and sometimes any bounded degree graphs, although we chose here to present the numerical results for degree 3 only. E.g., for problem MaxCut on 4 (resp 5) regular graphs we obtain bounds of 0.57 (resp. 0.56), with shorter annealing times. We applied this technique to MaxCut and MIS to show how it can be used to access the numerical values of the approximation ratios for cubic graphs. We then discussed the non-tightness of the Lieb-Robinson bound using MaxCut and drew some "worst" cases graphs to argue that QA can probably reach an approximation ratio for this problem, significantly better than the one we can actually prove. 

%\subsection{Future work}

Let us suggest some further research directions. 
In this paper, we developed tools to retrieve locality in the standard formulation of adiabatic quantum computing and use this relaxed locality to deduce lower bounds on the performances of QA for MaxCut and MIS on cubic graphs. There are two main directions to improve upon these results. The first direction would be to refine the analysis of constant time QA using less restrictive locality arguments, as we started to do in Appendix~\ref{app:hints} for MaxCut.

%In this paper, we followed the general setting of adiabatic quantum computing, but quantum annealing allows more degrees of freedom to improve the efficiency of the algorithm. 
The other direction would be to improve the algorithm itself as quantum annealing allows more degrees of freedom to work on the efficiency. This improvement can take many shapes.
%Another parameter is the initial state. 

We started our QA with a simple initial state, corresponding to the uniform superposition of all bitstrings of length $n$. Nevertheless, given that the LR bound used here does not depend on this initial state, there may be more suitable choices yet to explore.  Similarly, to boost the results of approximation with QA, one can modify directly the Hamiltonians. We can choose to work with a different initial Hamiltonian $H_0$, e.g., the ones of~\cite{Childs2002} who deal with bit swaps, whereas here we used bit flips. We can also change the target Hamiltonian of the evolution without changing the objective function. For example, in MIS, one can make a quantum evolution without the Hamming weight part \cite{farhi2020quantum}. Indeed, the state of minimal energy remains the same and we keep looking at the expected value of the total problem Hamiltonian reached by the final state of this particular evolution. The approximation ratio slightly improves up to 0.35.

Another lead which has been suggested by the physics field of Shortcut to Adiabadicity (STA) \cite{Gu_ry_Odelin_2019} encourages the use of a perturbative Hamiltonian to accelerate the evolution toward the adiabatic, optimal behavior. For example, it seems that a perturbative Hamiltonian of the form $H_e = [H_0,H_1]$ is a good initial choice as it increases the approximation ratio and decreases the time at which it is obtained.  

Eventually, it is important to notice that one %we do not have to
should not compare directly QAOA and QA developed here because, unlike QAOA, we did not try to optimize several parameters and degrees of freedom available. 
If we allow arbitrary trajectory for QA, QAOA reduces to a particular case of QA. This raises the natural question of whether QA can beat QAOA for a particular choice of schedule.
%In fact, one can show that QAOA results can be reached by QA following a specific trajectory. Let us define QAOA($\beta, \gamma$)\marginpar{\ioan{OK... pour ceux qui connaissent bien QAOA et ses paramètres $\beta, \gamma$}} for $p=1$ and fixed parameters and QA($T,\alpha$) a quantum annealing with the following Hamiltonian $H(t)=\mathbb{1}_{t\leq \alpha T} H_1 + \mathbb{1}_{t> \alpha T} H_0$, called the "bang-bang" schedule, we have that $\max_{T,\alpha} QA(T,\alpha) = \max_{\beta,\gamma} QAOA(\beta,\gamma)$. 
Until \cite{Brady_2021}, QAOA trajectory, also called "bang-bang", was thought as the  optimal schedule from the optimal control theory point of view. We know now that a "bang-anneal-bang" schedule, meaning that the schedule is time continuous between the two bangs, can be better, but also harder to find. All this to say that there is still a wide search space to improve QA, and we strongly believe that these directions can be combined with our LR based technique in order to obtain better approximation ratios.

%\newpage
%%
%% Bibliography
%%

%% Please use bibtex, 

\bibliography{shortQA}
%\vfill

%\pagebreak
\section{Proof of Proposition~\ref{LR-bound}}
\label{sec:proof}

We note $U_I(t) = (U_{0,t}^\Omega)^\dagger U_{0,t}^G$, the evolution in the interaction picture of $V_I(t) = (U_{0,t}^\Omega)^\dagger V(t) U_{0,t}^\Omega$ where the perturbation $V$ is $V(t)=H_G(t)-H_\Omega(t)$
\begin{align*}
    \left \lVert (U_{0,t}^G)^\dagger O_X U_{0,t}^G -(U_{0,t}^\Omega)^ \dagger O_X U_{0,t}^\Omega \right \rVert
    &=\left \lVert \int_0^t ds \frac{d}{ds} \left( U_I(s)^\dagger (U_{0,t}^\Omega)^\dagger O_X U_{0,t}^\Omega U_I(s) \right) \right \rVert \\
    &=\left \lVert \int_0^t ds  \left( U_I(s)^\dagger \left [(U_{0,t}^\Omega)^\dagger O_X U_{0,t}^\Omega, V_I(s) \right] U_I(s) \right) \right \rVert \\
    &=\left \lVert \int_0^t ds  \left( (U^G_{0,s})^\dagger \left [(U_{s,t}^\Omega)^\dagger O_X U_{s,t}^\Omega, V(s) \right] U^G_{0,s} \right) \right \rVert \\
    &=\left \lVert \int_0^t ds  \frac{s}{T} \left( (U^G_{0,s})^\dagger \left [(U_{s,t}^\Omega)^\dagger O_X U_{s,t}^\Omega, H_{1,\partial \Omega} \right] U^G_{0,s} \right) \right \rVert 
\end{align*}

The last equality is true because in $V$ we have every term in $\bar{\Omega}$, all the interactions terms between $\Omega$ and $\bar{\Omega}$ and the terms between two nodes of $\Omega$ that are not considered in $ \Omega$. The only interactions terms in $\partial \Omega$ lie in $H_1$ and we write it $H_{1,\partial \Omega}$. Thus, $V(s) = H_G(s) - H_\Omega(s) = H_{\bar{\Omega}} + \frac{s}{T} H_{1,\partial \Omega}$. Because $X \subset \Omega$, the left term in the commutator is strictly supported by $\Omega$, so the term with $H_{\bar{\Omega}}$ cancels out and we are left with the last line. The factor $\frac{s}{T}$ comes from the linear interpolation we are working with but it can be any chosen schedule. The final result uses the triangular inequality with the norm and integral and that $U_{0,s}^G$ is a unitary operator.

\section{Detailed intuitions}\label{app:hints}

Here we detail the intuition on QA evolution apply to MaxCut. The ratio obtained is not to be compared to the ones presented in this paper because we do not restrict the analysis of QA as a local algorithm anymore. We still investigate what QA can expect to reach in a short constant runtime.

\begin{figure}[ht]%
 \centering
 \subfloat[]{

 \begin{tikzpicture}
  \node[circle, draw, inner sep=1pt](v1) at (-1,0){1};
  \node[circle, draw, inner sep=1pt](v2) at (-1,1){2};
  \node[circle, draw, inner sep=1pt](v3) at (-1,-1){3};
  
  \node[circle, draw, inner sep=1pt](v4) at (1,0){4};
  \node[circle, draw, inner sep=1pt](v5) at (1,1){5};
  \node[circle, draw, inner sep=1pt](v6) at (1,-1){6};

  \draw (v1) -- (v2) -- (v5) -- (v4) -- (v1) -- (v3) -- (v6) -- (v4);
  \draw (v3) -- (v5);
  \draw (v2) -- (v6);
  \end{tikzpicture}\label{fig:a}}
 \subfloat[]{
 \scalebox{0.8}{
 \begin{tikzpicture}
  \node[circle, draw, inner sep=1pt](v1) at (-1,0){1};
  \node[circle, draw, inner sep=1pt](v2) at (-2,2){2};
  \node[circle, draw, inner sep=1pt](v3) at (-2,-2){3};
  \node[circle, draw, inner sep=1pt](v4) at (-1,3){4};
  \node[circle, draw, inner sep=1pt](v5) at (-1,1){5};
  \node[circle, draw, inner sep=1pt](v6) at (-1,-1){6};
  \node[circle, draw, inner sep=1pt](v7) at (-1,-3){7};
  
  \node[circle, draw, inner sep=1pt](v8) at (1,0){8};
  \node[circle, draw, inner sep=1pt](v9) at (2,2){9};
  \node[circle, draw, inner sep=1pt](v10) at (2,-2){10};
  \node[circle, draw, inner sep=1pt](v11) at (1,3){11};
  \node[circle, draw, inner sep=1pt](v12) at (1,1){12};
  \node[circle, draw, inner sep=1pt](v13) at (1,-1){13};
  \node[circle, draw, inner sep=1pt](v14) at (1,-3){14};

  \draw (v1) -- (v8) -- (v9) -- (v11) -- (v4) -- (v2) -- (v1) -- (v3) -- (v7) -- (v14) -- (v10) -- (v8) ; 
  \draw (v3) -- (v6) -- (v13) -- (v5) -- (v12) -- (v7) ;
  \draw (v4) -- (v14) ; \draw (v6) -- (v11) ; 
  \draw (v2) -- (v5); \draw (v9) -- (v12); \draw (v10) -- (v13);
  \end{tikzpicture}}\label{fig:b}}%
 \subfloat[]{
  \scalebox{0.7}{
 \begin{tikzpicture}
  \node[circle, draw, inner sep=1pt](v1) at (-1,0){1};
  \node[circle, draw, inner sep=1pt](v2) at (-2,2){2};
  \node[circle, draw, inner sep=1pt](v3) at (-2,-2){3};
  \node[circle, draw, inner sep=1pt](v4) at (-2,3){4};
  \node[circle, draw, inner sep=1pt](v5) at (-3,2){5};
  \node[circle, draw, inner sep=1pt](v6) at (-3,-2){6};
  \node[circle, draw, inner sep=1pt](v7) at (-2,-3){7};
  \node[circle, draw, inner sep=1pt](v8) at (-1,4){8};
  \node[circle, draw, inner sep=1pt](v9) at (-2.5,4){9};
  \node[circle, draw, inner sep=1pt](v10) at (-4,2.1){10};
  \node[circle, draw, inner sep=1pt](v11) at (-4,1){11};
  \node[circle, draw, inner sep=1pt](v12) at (-4,-1){12};
  \node[circle, draw, inner sep=1pt](v13) at (-4,-2){13};
  \node[circle, draw, inner sep=1pt](v14) at (-3.2,-4){14};
  \node[circle, draw, inner sep=1pt](v15) at (-1,-4){15};

  \node[circle, draw, inner sep=1pt](v16) at (1,0){16};
  \node[circle, draw, inner sep=1pt](v17) at (2,2){17};
  \node[circle, draw, inner sep=1pt](v18) at (2,-2){18};
  \node[circle, draw, inner sep=1pt](v19) at (2,3){19};
  \node[circle, draw, inner sep=1pt](v20) at (3,2){20};
  \node[circle, draw, inner sep=1pt](v21) at (3,-2){21};
  \node[circle, draw, inner sep=1pt](v22) at (2,-3){22};
  \node[circle, draw, inner sep=1pt](v23) at (1,4){23};
  \node[circle, draw, inner sep=1pt](v24) at (3.2,4){24};
  \node[circle, draw, inner sep=1pt](v25) at (4,2){25};
  \node[circle, draw, inner sep=1pt](v26) at (4,1){26};
  \node[circle, draw, inner sep=1pt](v27) at (4,-1){27};
  \node[circle, draw, inner sep=1pt](v28) at (4,-2.1){28};
  \node[circle, draw, inner sep=1pt](v29) at (2.5,-4){29};
  \node[circle, draw, inner sep=1pt](v30) at (1,-4){30};
  
  \draw (v1) -- (v2) -- (v4) -- (v8) -- (v23) -- (v19) -- (v17) -- (v16) -- (v1) -- (v3) -- (v7) -- (v15) -- (v30) -- (v22) -- (v18) -- (v16);
  \draw (v2) -- (v5) -- (v10) -- (v29) -- (v22);
  \draw (v5) -- (v11) -- (v26) -- (v20) -- (v25) -- (v13) -- (v6) -- (v12) -- (v27) -- (v11);
  \draw (v8) -- (v30); \draw (v15) -- (v26); \draw (v12) -- (v23); \draw (v3) -- (v6); \draw (v17) -- (v20); \draw (v21) -- (v27); \draw (v29) -- (v13); 
  \draw (v18) -- (v21) -- (v28) -- (v14) -- (v7);
  \draw (v4) -- (v9) -- (v25); \draw (v19) -- (v24) -- (v10); \draw (v9) -- (v28); \draw (v14) -- (v24);
  \end{tikzpicture}}\label{fig:c}}%
 \caption{Worst graphs from a QA point of view in short running time. (a) $G_6$ with minimum cycle size $=4$. (b) $G_{14}$ with minimum cycle size $=6$. (c) $G_{30}$ with minimum cycle size $=8$. Each edge has exactly the same neighborhood.}%
 \label{fig:worst}%
\end{figure}
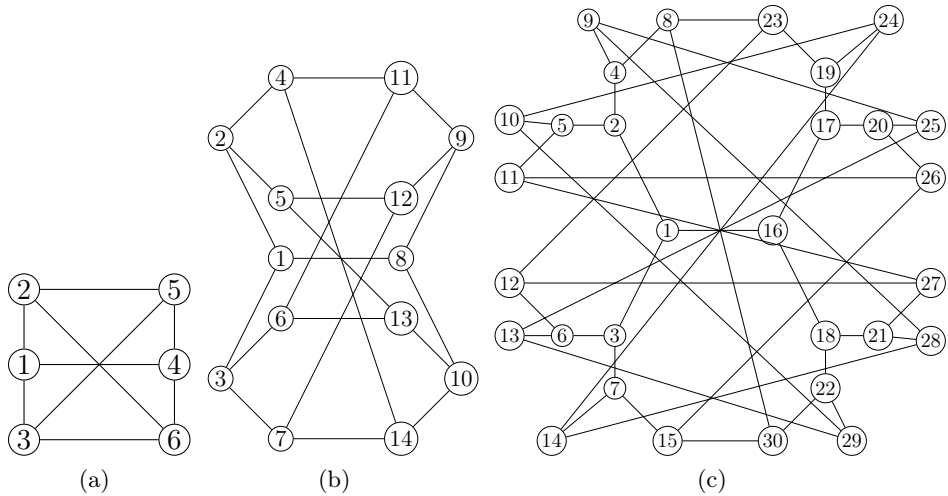

With more informal reasoning, we think that short-constant-runtime QA can reach an approximation ratio of 0.73. One can try to conduct a similar reasoning as the one hinted for the cycle in Appendix \ref{sec:cycle} which presents an easy and very illustrative case to grasp the notion of locality in short runtime. This latter case drives us to think that studying only small graphs is enough to have a clear idea of the approximation ratio for small $T$. 
We decide to focus on bipartite, edge-transitive graphs, i.e. the graphs have exactly the same neighborhood at any distance for each edge. So we look at different neighborhoods at different distances and experimentally select the "worst" one. Then, we construct graphs such that each edge has this specific neighborhoods for different distances. Intuitively speaking, we note that adding a cycle helps because it reduces the number of nodes lying at a given distance from an edge.
For similar reasons, configurations obtained from $\Omega_1$ or $\Omega_2$ do not seem critical for the approximation ratio. And because we are working with small enough graphs (at small enough $T$), odd cycles create a non negligible offset at the beginning, i.e. the random guess is already strictly better than 0.5.
In this way, we construct 3-regular graphs where each edge have a neighborhood similar to a double binary tree of depth 1,2 and 3. Indeed, experimentally, these neighborhoods produce the slowest convergence of the middle edge's energy toward its optimal value (1 in the case of bipartite graph). These graphs are denoted $G_6$, $G_{14}$ and $G_{30}$, the index corresponding to the number of their nodes (Fig. \ref{fig:worst}). 

%Finally, Figure~\ref{fig:worst} depicts the cubic graphs in which each edge has a ball at distance 1 of type $\Omega_3$, by increasing size. , we believe these are the worst examples for QA in very short time.

\begin{figure}
    \centering
    \includegraphics[scale=0.6]{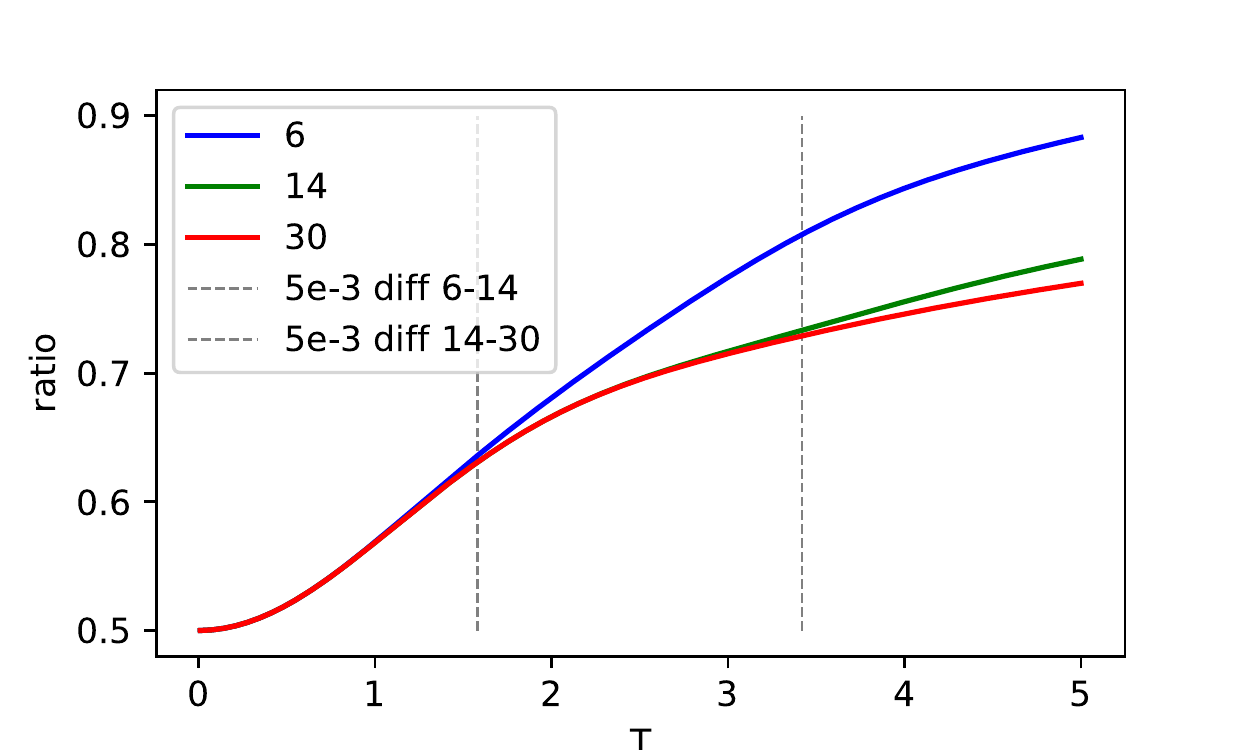}
    \caption{Ratio reached by $G_6$ (blue line), $G_{14}$ (orange line) and $G_{30}$ (red) against running time $T$. The black dotted lines shows when there is a difference of at least 0.005.}
    \label{fig:ratio}
\end{figure}

In Fig. \ref{fig:ratio}, we plot the approximation ratio reached by these graphs as a function of $T$. Until the first dotted line, all 3-regular graphs up to 20 vertices that we tested have a ratio that is above or equal to the blue and orange curves. Then between the two dotted lines, we have the first split, QA had enough time to make the difference between cycles of size 4 and larger cycles. But still, until the second dotted line, all tested graphs have a ratio greater or equal than the orange curve (or red). Then, we have the second split, the quantum system can access information that is at distance 3 in the graph and thus can make the difference between $G_{14}$ and $G_{30}$.

From these observations, we conclude that QA with linear interpolation probably reaches an approximation ratio for MaxCut on 3-regular graphs of 0.73 (minus an arbitrary $\varepsilon$) at $T_1=3.47$ and, moreover, from 0 to $T_1$, the approximation ratio is described by the orange curve. This value is the same as the middle edge energy of $\Omega_3$ minus 0.005 at $T_1$.

\section{Cycle case}
\label{sec:cycle}

\begin{figure}[ht]
    \centering
    \includegraphics[scale=0.4]{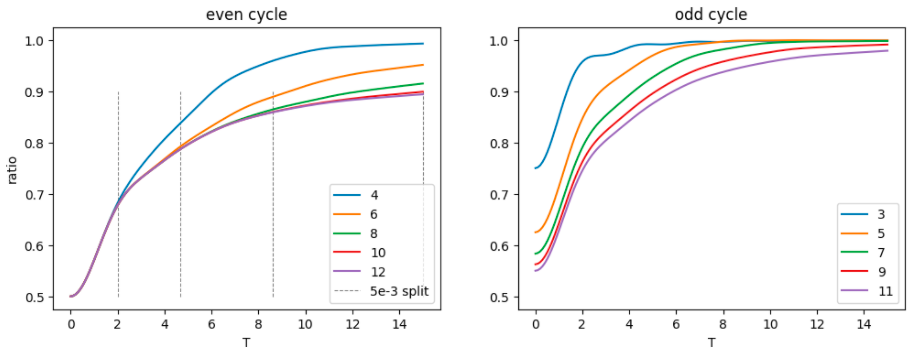}
    \caption{Approximation ratio reached by even cycles (left) and odd cycles (right). The dotted lines indicate the $5.10^{-3}$ splitting.}
    \label{fig:cycles}
\end{figure}

The 2-regular graphs represented by cycles is a very illustrative case to get the intuition of the locality in QA. It also clarifies the construction of the hardest cubic graphs (bipartite graphs with large even cycles). In Fig. \ref{fig:cycles}, we plot the approximation ratio against the runtime $T$ for even cycles (left) and odd cycles (right). The first observation is that odd cycles converge faster to 1 than even cycles and they start with an offset at $T=0$. Indeed, they do not have a clean cut so the random guess for a cut gives a ratio above 0.5 starting the evolution at $0.5 \frac{n}{n-1}$. We understand that having odd cycles helps QA to decide faster. 

\begin{remark}
Because cycles are edge transitives, the ratio plot here is also the expectation value of the observable on one edge.
\end{remark}

\begin{figure}[ht]
    \centering
    \includegraphics[scale=0.3]{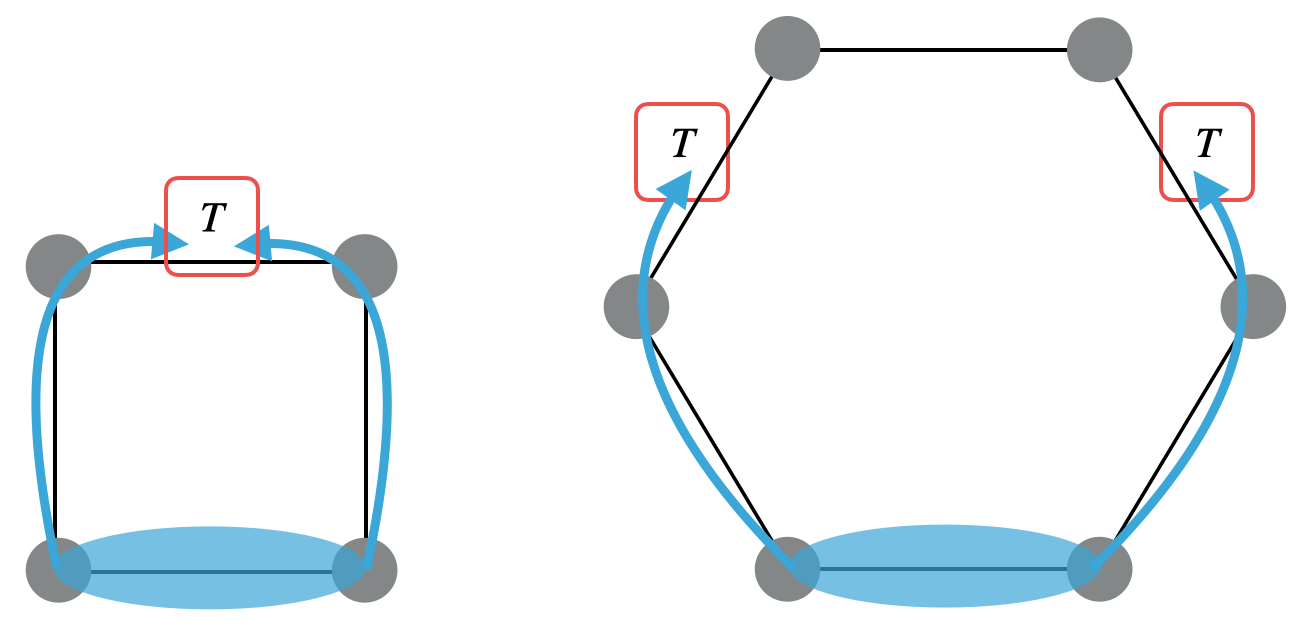}
    \caption{Information travel of the observable in shaded blue during time $T$. After this time, QA can make the difference between these two cycles.}
    \label{fig:cycle_travel}
\end{figure}

Looking closely at the even cycles, we see that until some time $T$ all ratios follow the same tract. At some point, the smallest cycle splits to go toward 1 faster, the others keep the same trend. Then this splitting scheme repeats again with the left cycles. The dotted lines indicate when there is more than $0.005$ difference between cycles of size $2n$ and $2n+2$. This conveys the idea that up to some time, QA does not have time to see far in the graph to make the decision. In this case, we can even tell "illustratively" how deep the information travels in the graph like in Fig. \ref{fig:cycle_travel}.

\end{document}